\documentclass[natbib,smallcondensed]{svjour3}     
\usepackage{graphicx}
\usepackage{booktabs}  
\usepackage{xspace}
\usepackage{lscape}
\usepackage{bm}
\usepackage{soul}
\usepackage{longtable}

\usepackage{xcolor}                       
\usepackage[colorlinks=true,citecolor=blue,urlcolor=blue,breaklinks]{hyperref}

\journalname{Reviews of Modern Plasma Physics}

\newcommand{\del}{\mbox{\boldmath $\nabla$}}
\def\vec{\boldsymbol}
\def\avg{\bar}

\usepackage{color}

\def\bl{Babcock--Leighton}
\def\mc{meridional circulation}
\def\mf{meridional flow}

\newcommand{\Fig}[1]{Figure~\ref{#1}}
\newcommand{\Figs}[2]{Figures~\ref{#1} and \ref{#2}}

\newcommand{\Eq}[1]{Equation~(\ref{#1})}

\newcommand{\Sec}[1]{Section~\ref{#1}}

\newcommand{\Tab}[1]{Table~\ref{#1}}

\def\avg{\overline}

\newcommand{\mps}{m~s$^{-1}$}

\begin{document}

\title{Solar Cycle Prediction: Challenges, Progress, and Future Perspectives}

\author{Bidya Binay Karak}

\institute{B. B. Karak \at 
              Department of Physics, Indian Institute of Technology (Banaras Hindu University), Varanasi, India \\
              \email{karak.phy@iitbhu.ac.in}           
}
\date{Received: date / Accepted: date}

\maketitle

\begin{abstract} 
Reliable prediction of the solar cycle is a formidable challenge, yet it is increasingly vital in our technology-dependent society as solar activity drives space weather.  Various methods, including precursors, nonlinear curve fitting and extrapolation, statistical and Machine Learning (ML) models, and dynamo and surface flux transport (SFT) models, were implemented to predict past cycles.  Analysing about 100 predictions for Solar Cycle 24 and over 130 for Solar Cycle 25, we find that most methods largely failed to predict the peak correctly: Cycle 24 was statistically predicted to be a strong cycle, whereas Cycle 25 was predicted to be a weak cycle. By and large, predictions made only after the cycle began became closer to reality. ML-based models also produced discouraging results. The polar field and its proxy-based predictions are the most physically supported approach to prediction; however, applying them much earlier, before the solar minimum, may yield inaccurate results.  Dynamo models are progressively improving both in understanding and in forecasting; however, they need to improve by accurately assimilating the observed polar field data and additional physics, such as meridional flow variations. Solar dynamo theory, complemented by the SFT model and observations, demonstrates that the prediction of a cycle before the time of its previous cycle's maximum is meaningless.   The current solar cycle is declining, and the community is now preparing for the prediction of the next cycle. Thus, this review will guide future studies.
\end{abstract}

\keywords{Solar physics \and Solar Activity \and Solar cycle \and Solar dynamo}

\setcounter{tocdepth}{3}
\tableofcontents



\section{Introduction}
\label{section:introduction}

One of the intriguing cycles of nature is the solar cycle, which is the cyclic variation of the magnetic field \citep{babcock59}. The magnetic field strength of the Sun is observed to oscillate with a period of 11 years, and in every cycle the polarity of the field also flips \citep{Hat15}. The strong magnetic field, which is produced in the interior, erupts through the surface in the form of concentrated bipolar magnetic regions (BMRs), the white-light (or intensity continuum) counterpart of these regions being popularly known as sunspots or active regions \citep{LG2015}.  The number of sunspots, for which we have fairly systematic records dating back to approximately 1610, also varies over a period of about 11 years, making the solar cycle popularly known as the sunspot cycle \citep{Schwabe1844}.  It is the magnetic field of the sunspot and decayed active regions, which is quite dynamic and complex, that usually produces energetic events like coronal mass ejections (CMEs) and flares, through which huge amounts of charged particles, radiation, and magnetic field are ejected from the Sun \citep{Chen11, Cliver22, GEORGOULIS2024}. These short-wavelength radiation, energetic particles and magnetic field disturb the space environments around the Sun, including the space around the Earth and other planets, and drive the so-called space weather \citep{Temmer21}. Besides producing the beautiful aurora, space weather causes hazardous effects on us, including (but not limited to) disrupting satellite-based communications, satellite failure, disturbance in GPS communication, power grid failure, impacting astronauts in space, and disturbance in long-distance radio communication \citep{Schrijver15, IPOTSE408340}.  In June 2013, it was estimated that if a similar Carrington event \citep[a strong space weather event that occurred on Sept 1--2, 1859;][]{CD13} happens at present, then the cost of damage in the US alone can be as high as \$0.6--2.6 trillion \citep{Schrijver15b}, emphasising the impacts of space weather effects \citep{Nandy25}.

\begin{figure}
\centering
\includegraphics[width=0.75\columnwidth]{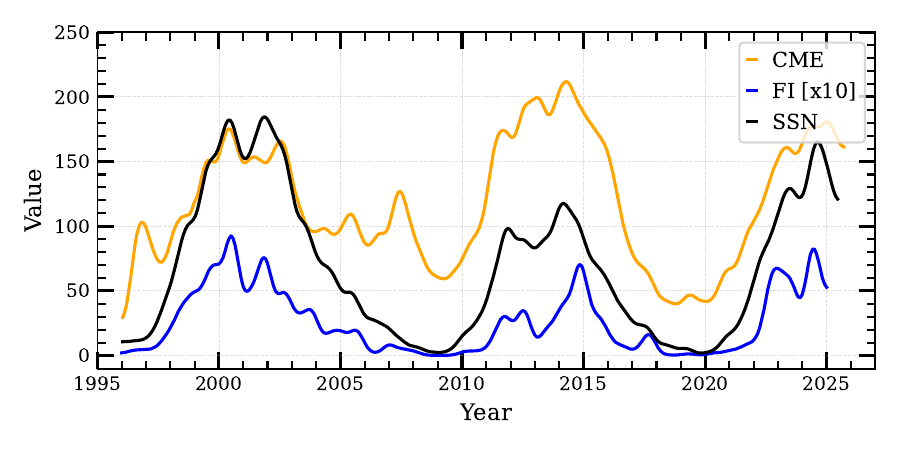}
\caption{
Variation of the monthly sunspot number (ISSN V2.0; black curve), flare index (blue), and monthly number of coronal mass ejections (SOHO/LASCO CME; orange) as functions of time. 
}
\label{fig:SSN_CME_Flare}
\end{figure}

As all these energetic space-weather candidates (CMEs and flares) are produced from the sunspots and decayed active regions, and the number of sunspots has a cyclic variation, we observe a cycle variation of the CMEs and flares in congruence with the sunspot cycle \citep{Scott15}. \Fig{fig:SSN_CME_Flare} shows the variation of the flares and CMEs 
overplotted with the number of sunspots for the last two and a half cycles. While there are large fluctuations on a short time scale of about a year, in general we observe that near the solar maximum, when the Sun produces the highest number of sunspots, the numbers of flares and CMEs are also generally high. The geomagnetic disturbances, as quantified by the aa-index variation, also showed a good correlation with the solar cycle \citep{chapman20, Hajra21}. These observations imply that solar maxima are the times when we expect more disturbances in space and vice versa. Hence, the condition of the space around the Sun goes through quiet to violent phases in a cyclic manner with a period of 11~years, coinciding with the sunspot cycle.  

\begin{figure}
\centering
\includegraphics[width=1.05\columnwidth]{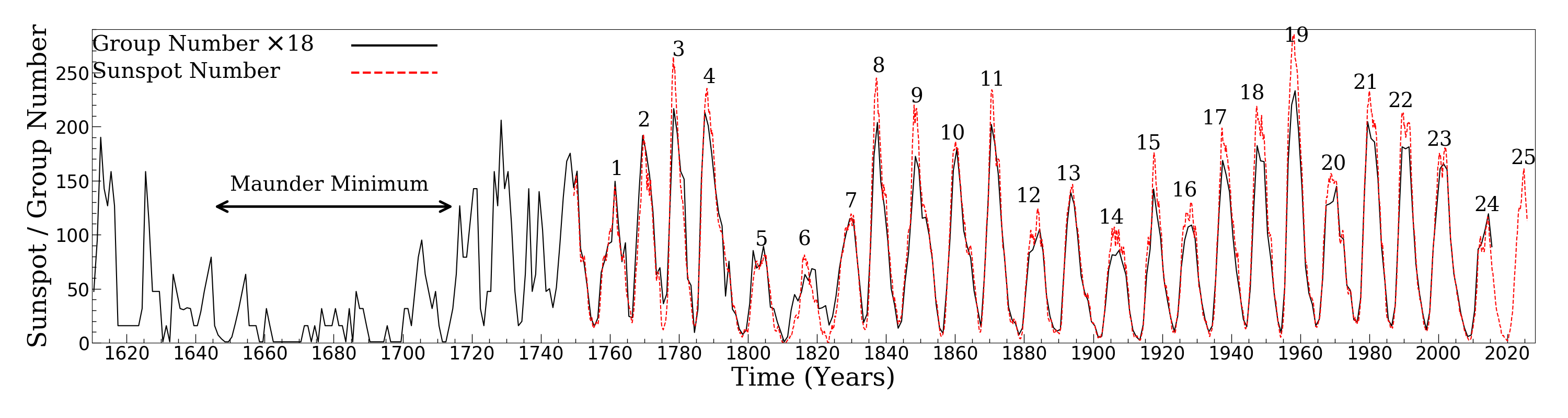}
\caption{
Solid/black: Yearly variation of the group sunspot number.  Dashed/red: Monthly smooth sunspot number (ISSN V2.0). Cycle numbers starting from Cycle~1 are marked near their peaks.
}\label{fig:obs_ssn}
\end{figure}

However, there is more to this cyclic (predictable) picture, and that becomes apparent when we observe the solar cycle on a long time scale. In \Fig{fig:obs_ssn}, we observe that the solar cycle amplitude had a fairly large variation in the past. There was a time when the Sun produced strong cycles (e.g., during cycles 17--19, called the Modern maximum), and there were times when the Sun produced weak cycles, e.g., during cycles 4--5 (so-called Dalton minima) and even produced Maunder minimum (during 1645 -- 1715) when the Sun showed only a few spots for several years \citep{Usoskin2017, Hayakawa25}. While there is a trend, as indicated by an envelope of the so-called Gleissberg cycle with a period of approximately 98 years, the amplitude of the solar cycle varied in quite an irregular fashion \citep{Biswas23}. Extreme space-weather events are more frequent in strong cycles \citep{Owens21, Kar26}. These indicate that the level of space weather disturbance (as caused by flares and CMEs) also varies irregularly, leading to a slow, long-term modulation of the space weather condition, also known as the space climate \citep{MUM07, Lockwood18}.         

As mentioned earlier, the prediction of the space weather condition well in advance is useful for planning and preparation of future space missions and also to determine the lifetime of satellites in low-Earth orbits \citep{OZ19, Miteva23, Oliveira24, Baruah}. In particular, if we can predict the solar activity level and thus the space climate conditions, say for the next decade, then this is important data for space organisations and satellite companies. Prediction is also useful for constraining the mechanism of the generation of solar magnetic field---the solar dynamo for which various potential models are available but no consensus \citep{Cha20, Kar23}.  However, as seen in \Fig{fig:obs_ssn}, the prediction of the strength of the solar cycle is challenging. The attempt for cycle prediction began at least about 50 years ago when researchers began predicting the amplitude of solar cycle 23 \citep[][and references therein]{Sch78, Joselyn97}. With increasing understandings of the mechanism of the solar cycle and the availability of observed data, solar cycle prediction has become a major branch of solar physics, producing more than 100 research publications just for the prediction of Cycle 24 and about 150 predictions for Cycle 25 according to the Astrophysics Data System \citep{Petrovay10, Pesnell12, Petrovay20, Nandy21, Bhowmik23, Miesch25}. Existing prediction methods can be broadly categorised into three groups.  These include (i) physics-based predictions, employing the underlying physical mechanism working behind the formation of the solar cycle, the solar dynamo, (ii) precursor-based predictions, utilising the feature(s) of the solar cycle that carries (carry) the information of the future, (iii) extrapolation methods, utlizing the pattern of the past cycles and employing some (purely mathematical and Machine Learning models) analysis of the solar activity time series. Keeping in mind that there are excellent review articles \citep{Petrovay10, Pesnell12, Petrovay20, Nandy21, Miesch25} detailing the prediction methods of solar cycles, here we shall pay special emphasis to the categories (i) and (ii) after reviewing the statistics of the predictions made using all methods.  We shall highlight predictions from the past two solar cycles and how they helped improve predictions for future cycles.

\begin{figure}
    \centering
    \includegraphics[width=0.9\linewidth, angle = 0]{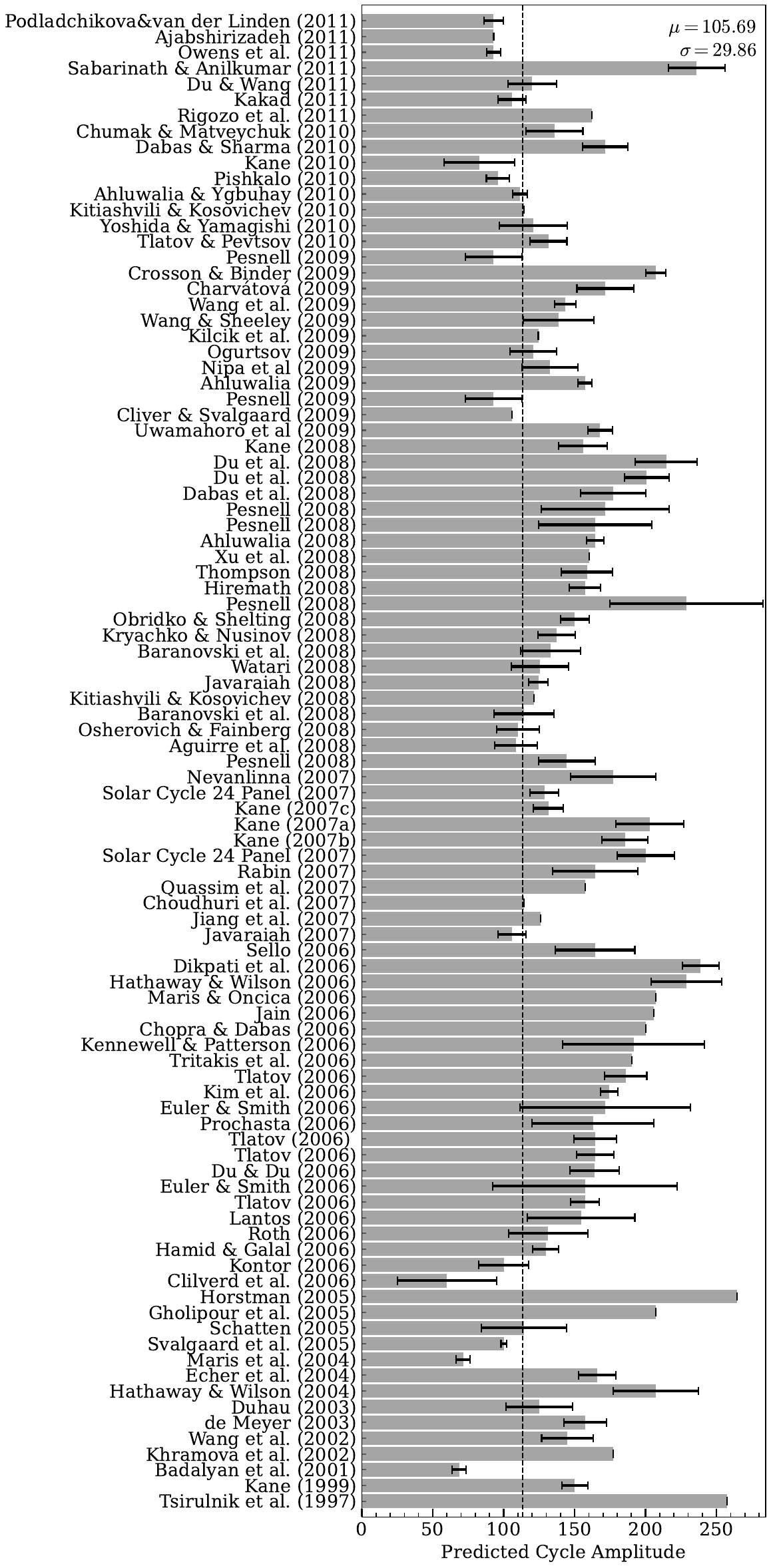}
    \caption{Predicted values of the peak sunspot number for Solar Cycle 24 made by  different authors are shown in chronological order. The dashed line marks the observed peak (113.3).
}
    \label{fig:sc24}
\end{figure}

\begin{figure}
    \centering
    \includegraphics[width=1.2\linewidth, angle = 0]{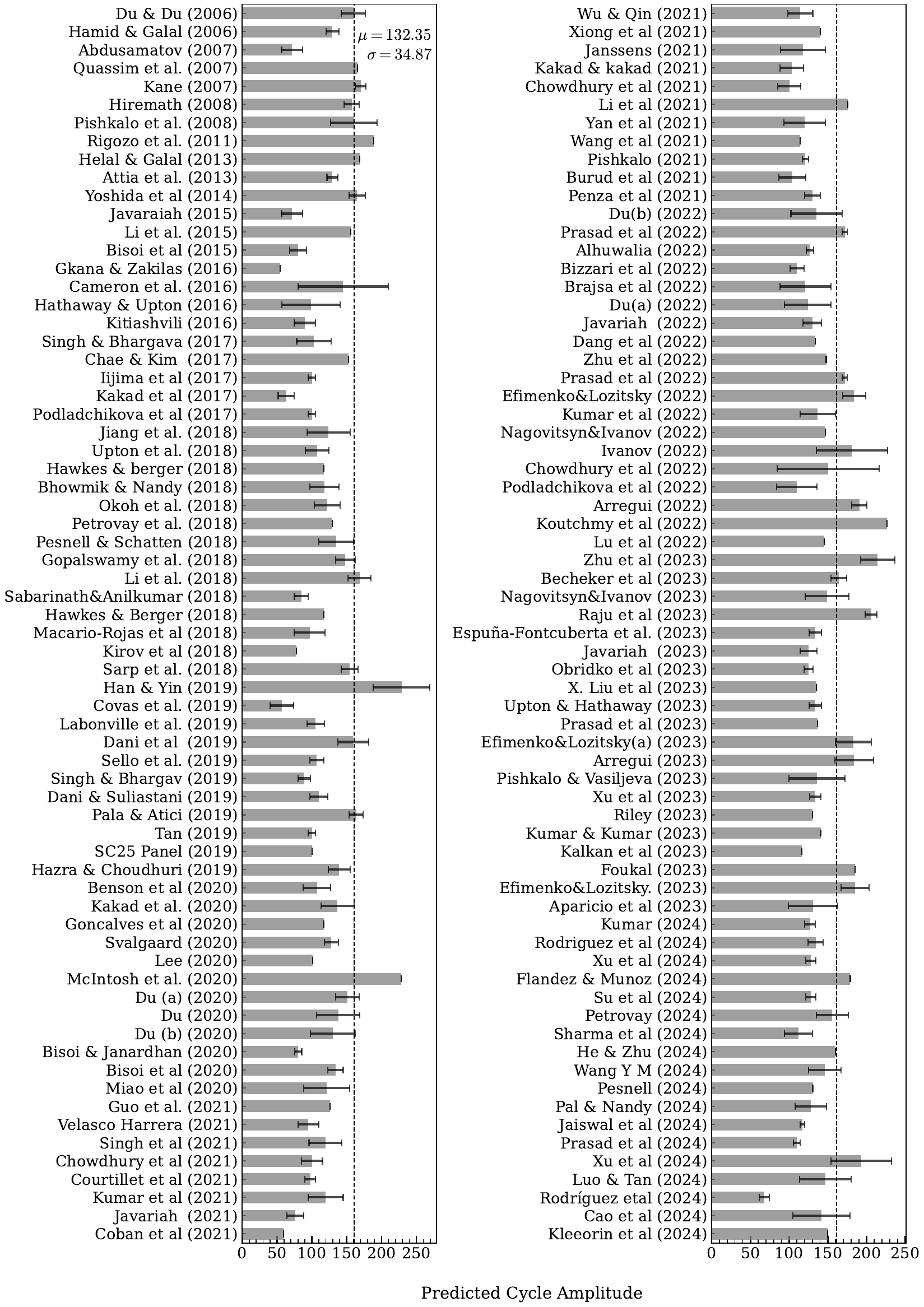}
    \caption{The same as \Fig{fig:sc24} but for Cycle~25. The dashed line points to the observed peak sunspot number at 161 (13-month smoothed value). The details of the references are listed in \Tab{tab:cyc25}.
    }
    \label{fig:sc25}
\end{figure}

\section{Summary of past predictions: Emphasising the challenges}

In \Figs{fig:sc24}{fig:sc25}, we display the predicted sunspot numbers of solar Cycles 24 and 25 made by various authors in chronological order. As most of the predicted values for Cycle 24 were repeatedly referred in previous reviews \citep{Pesnell12, Nandy21}, they are not cited separately here, although some new predictions are included here which were not cited in previous reviews. However, for Cycle 25, several new predictions have been added to the list and some values have also been corrected. Therefore, for completeness, all predictions made for Cycle 25 up to 2024 are listed in \Tab{tab:cyc25}. The predictions after 2024 are ignored here because by that time it was very clear about the cycle peak based on the increase rate of the cycle.  The following inferences can be made from these data.
 
For solar Cycle 24 (\Fig{fig:sc24}):\\
(i) No consensus on the predicted values. \\
(ii) The range of the predictions is quite large, going from less than 50 to as high as 260.\\
(iii) The mean ($\mu$) and the standard deviation ($\sigma$) are 105.88 and 29.97, respectively. Thus, the observed value\footnote{Several cycles, including Cycle 24,  display multiple peaks \citep{KMB18}. Therefore, assigning a number to the peak of a cycle is subjective, but we take the highest peak value as obtained from the 13-month smoothed ISSN V2.0 data, which is also commonly used in the literature to measure the strength of the cycle. } (113.3) significantly deviates from the mean. \\
(iv) 
The predictions made in 2009, the end of the cycle, are in no way superior to those made earlier, implying that even waiting a few more years did not improve the predictions much. 

Now, for Cycle 25 (\Fig{fig:sc25}):\\
(i) Again, there is no good news. Although the number of predictions has increased considerably, there is no consensus on the predicted values. The mean (132.20) is significantly below the observed value (161). \\
(iii) There is not much improvement in the prediction until the start of the cycle. Only after 2022 (when the cycle has clearly started growing), the values tend to become closer to reality. Even in 2024, when the cycle's growth rate could reasonably indicate its peak, several prediction methods produced significant deviations. \\
(iv) Except for the predictions made in 2006--2007, almost all the predictions during 2015--2021 were below the observed ones. 

\begin{figure}
    \centering
    \includegraphics[width=0.495\linewidth]{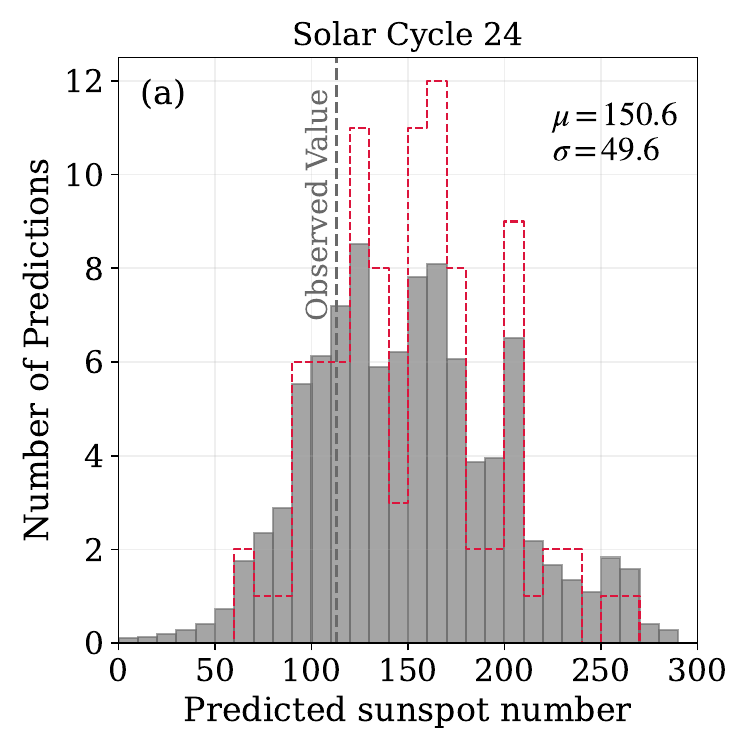}
    \includegraphics[width=0.495\linewidth]{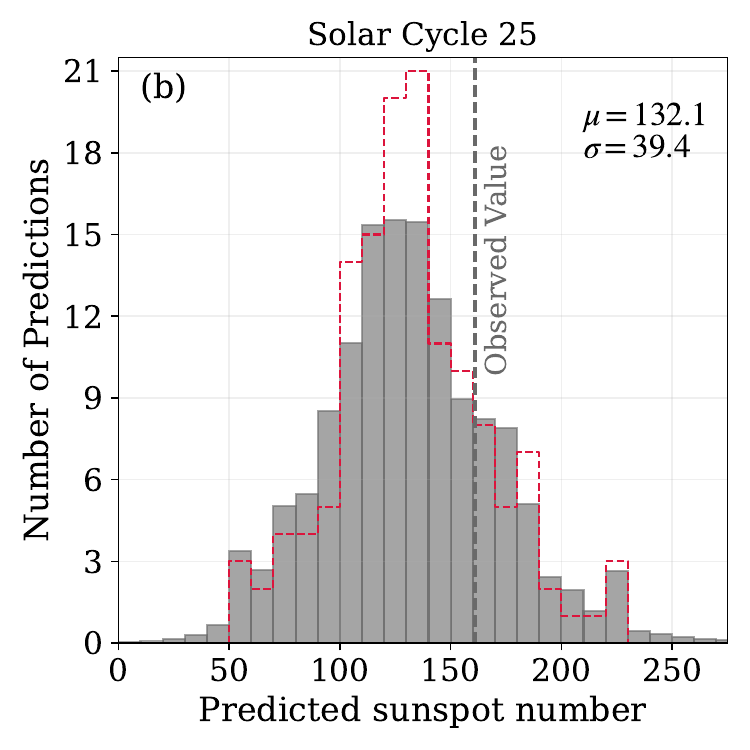}
    \caption{
Distributions (with bin size of 10) of the predicted peak sunspot numbers for the solar Cycles 24 (left) and 25 (right). The red/dashed distributions are obtained by ignoring the error of the prediction, while the filled/grey ones are obtained by considering the error (see text for more details).  Here, as well as for the previous two figures, predictions made on and before 2015 are scaled up by a factor of $1.43 \pm 0.1$ as given in \citet{Nandy21} to conform to the new version V2.0 from V1.0 \citep{Clette14}. The vertical lines represent the actual/observed values: 113 for Cycle 24 and 161 for Cycle 25.
}
    \label{fig:pred_dist}
\end{figure}

To obtain a clearer view of the predictive performance over the past two cycles, we plot the distributions of predicted sunspots. If we ignore the error ranges (which are not provided in all prediction publications), then the distribution is computed based on the occurrence frequency in each bin of size 10, which is shown by the red/dashed lines in \Fig{fig:pred_dist}. However, if we want to consider the error ranges, there are many ways to construct the distribution. The way we compute it is the following. For predictions whose error range is unavailable, we set the distribution frequency to 1 in the bin where the predicted sunspot number falls. However, for predictions having an error range, we obtain the frequency by assuming a Gaussian distribution with the predicted value as the mean and the error as $\sigma$. Grey/filled regions show this distribution in \Fig{fig:pred_dist}. 
From this figure, we see that Cycle 24 was statistically predicted to be a strong cycle, while Cycle 25 was predicted to be a weak cycle. 
In addition, the spread of the distribution of Cycle 25 is not reduced significantly compared to Cycle 24.  It is pretty frustrating that even after a decade of research, our understanding of solar cycle prediction has not improved, clearly demonstrating the difficulties in the problem.  
Now we shall move on to the physics of solar cycle and its predictability.

\section{The physical mechanism of the solar cycle}
\label{sec:physicalmodel}
To understand the physics of solar cycle prediction, we shall first discuss the physical mechanism governing the solar cycle, the dynamo. 
The solar cycle is governed by a process called the  dynamo, in which the kinetic energy of the plasma amplifies and maintains the magnetic energy against its diffusion through turbulent flow \citep{Mof78, Char23}. The differential motion (large-scale longitudinal flow) of the Sun generates a strong toroidal field by shearing a weak poloidal field in the solar convection zone. The toroidal field becomes intermittent, forming flux tube-like structures within the convection zone. 
These tubes become buoyant, rise up to the surface to erupt as sunspots (technically, the bipolar magnetic regions), which eventually decay and disperse to generate a poloidal field of opposite polarity (with respect to the existing field). This poloidal field first cancels the old field and continues to grow to form the polar field\footnote{The polar field is the radial component of the poloidal field measured near the polar regions.} for the next cycle, which eventually produces a toroidal field of opposite polarity. The toroidal field again produces sunspots and continues the cyclic process. While the fundamental mechanisms behind this cyclic dynamo process were presented in several historic publications \citep{Pa55, Parker75, Par79, Ba61, Le64, Leighton69}, researchers have offered various pieces of observational evidence and theoretical investigations over the past few decades, which, when combined, allow us to confirm the basic process underlying the solar cycle \citep{CS23}. Some of these key contributions and observations are listed below.

\begin{figure}
    \centering
    \includegraphics[width=0.75\linewidth]{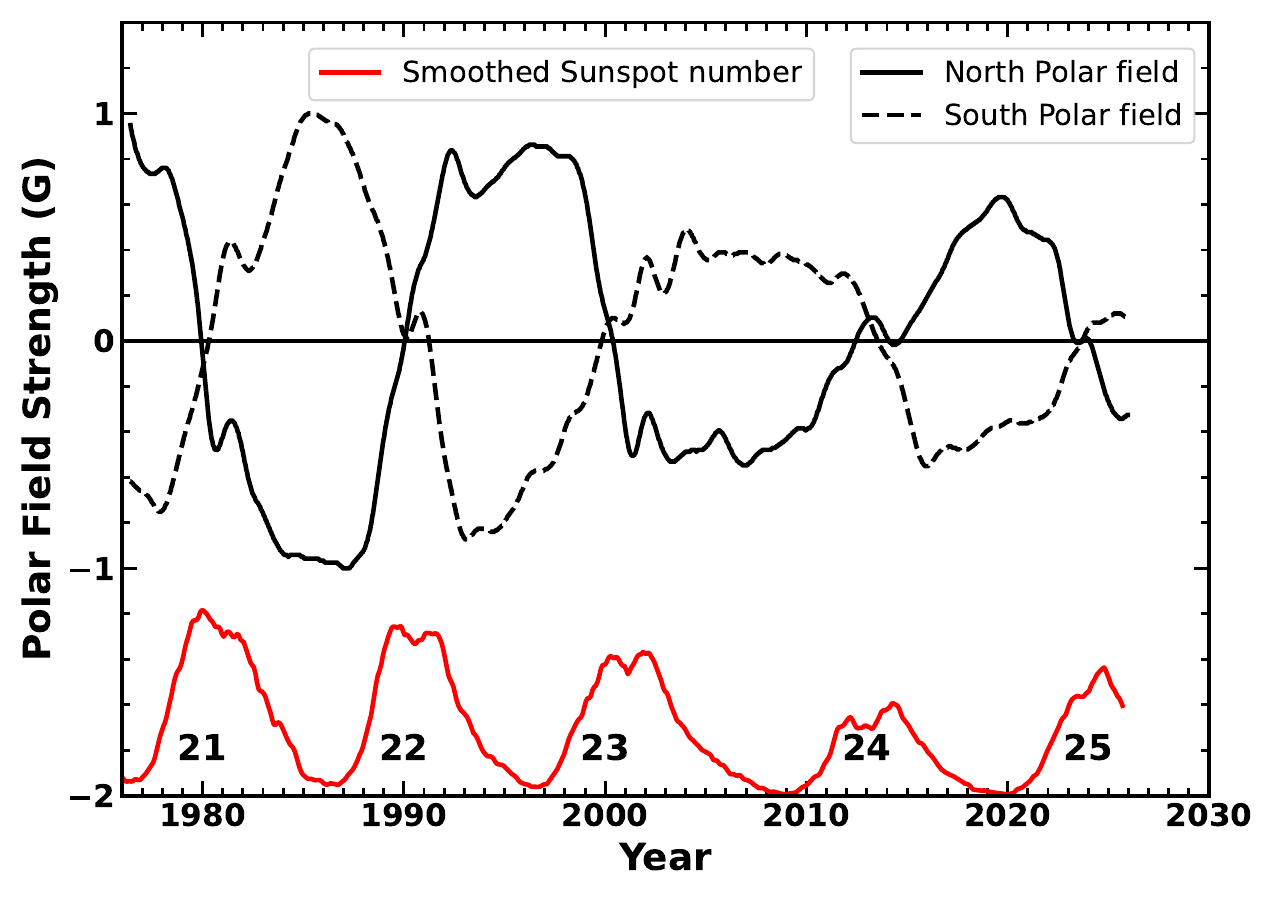}
    \caption{The variations of the polar field (top plots, a measure of the poloidal magnetic field of the Sun) and the  (13-month smoothed) sunspot number (bottom plot), a measure of the toroidal field strength. Note the polar field peaks around solar minima and reverses during solar maxima.}
    \label{fig:rad2ssn}
\end{figure}

\begin{figure}
    \centering
    \includegraphics[width=0.55\linewidth]{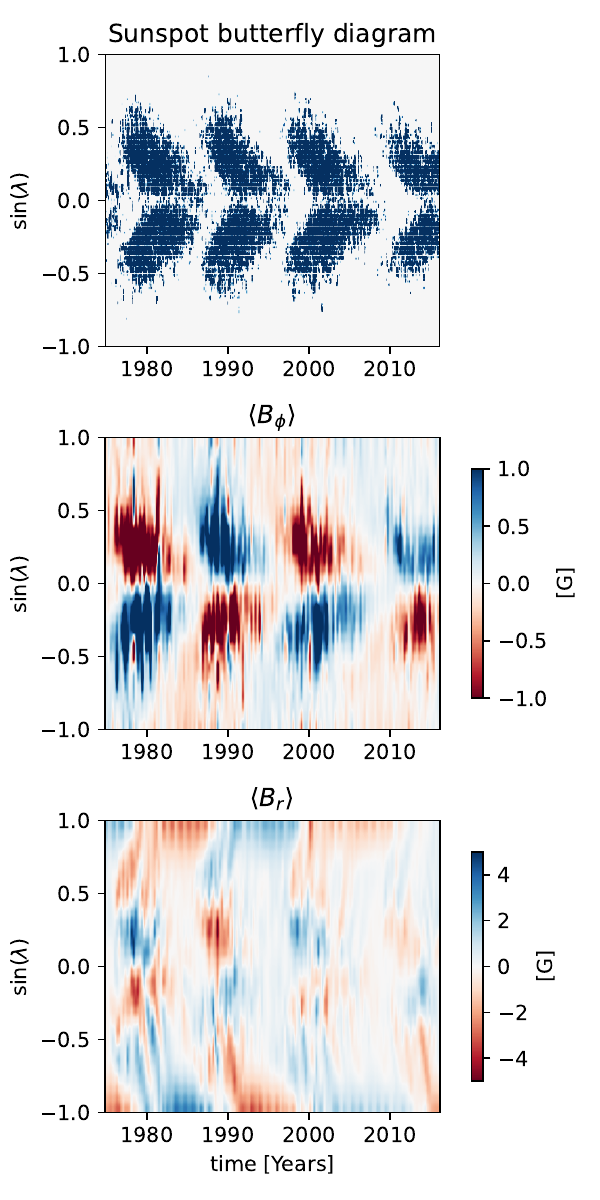}
    \caption{
    Top panel: Butterfly diagram of sunspot group obtained from Royal Greenwich Observatory(RGO)/NOAA.  Middle: Longitude-average azimuthal magnetic field (data source: Wilcox Solar Observatory, WSO).
 Bottom: Longitude-averaged radial field (data from WSO). 
 Note the proximity between the sunspot and the toroidal field, respectively, in the top two panels, and the phase lag of the poloidal field (bottom), becoming maximum when the sunspot/toroidal field goes to a minimum.  Figure reproduced from \citet{CS23} with permission from authors.
 }
    \label{fig:BrBphi}
\end{figure}

\begin{figure}
    \centering
    \includegraphics[width=0.75\linewidth]{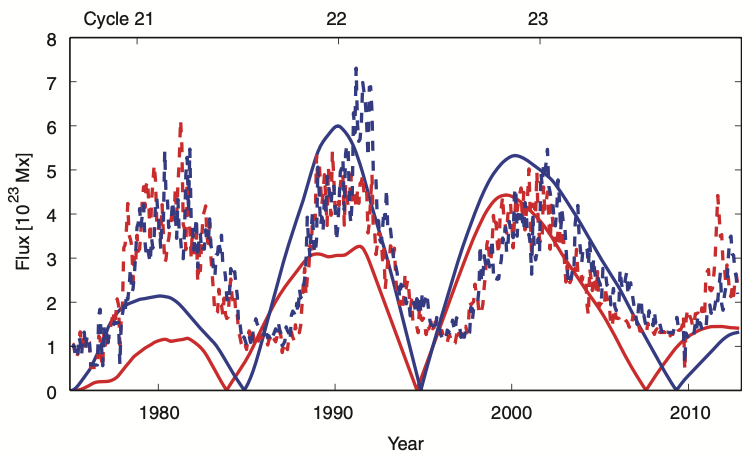}
    \caption{Dashed lines: The observed magnetic flux on the solar surface (red: north, blue: south) derived from Kitt Peak synoptic magnetograms. Solid lines: The computed net toroidal flux using the observed polar field and the differential rotation. Reproduced from \citet{CS15} with permission from authors.}
    \label{fig:cs15}
\end{figure}

(i) The observed polar magnetic field and sunspot number, which is produced from the toroidal field, are observed to oscillate out of phase as shown in \Fig{fig:rad2ssn}. In addition, the weak diffuse radial field and the toroidal field at low latitudes are measured, which for the last four cycles are shown in \Fig{fig:BrBphi}. In this figure, we observe that these two fields oscillate in a cyclic manner; the polar field is becoming strong at the solar minimum, whereas the toroidal field becomes prominent at the times of solar maxima. The polarity of these fields flips every 11-year cycle. 

(ii) The differential rotation in the entire SCZ has been known for the last several decades \citep{Schou98}. Thus, one can compute the toroidal flux generated from the observed polar field through shearing by the differential rotation \citep{CS15}, and this value is indeed in good agreement with the observed toroidal flux (\Fig{fig:cs15}). In addition, the observed polar flux \citep{CCJ07, Kumar21, Kumar22} and its proxies, such as the A(t)-index \citep{Makarov01}, polar faculae \citep{S91, Muno13}, and polar network counts \citep{Priy14, mishra25}, the geomagnetic aa-index \citep{WS09}, around the solar minimum, show a strong correlation with the amplitude of the next cycle. These suggest that the poloidal field is the source of the next sunspot cycle.  

(iii) The observed magnetic field in BMRs and active regions shows a general rule known as Hale's law \citep{Hale19, HN25}. According to this, the leading spots or poles of BMRs (located at higher longitudes) have the same magnetic polarity within a cycle and in a hemisphere. This polarity pattern alternates between hemispheres and cycles.
These feature of active regions/BMRs, complemented with the observed toroidal flux and its proximity to the observed latitudinal distribution of sunspots (\Fig{fig:BrBphi}), implies that the BMRs are the buoyant structures of the rising toroidal flux systems arising from the deep CZ. 

One mysterious feature of the solar cycle is its systematic increase in the tilt with the latitude--Joy's law \citep{Hale19}, for which the thin flux tube model, employing the effect of Coriolis force on the rising magnetic flux tube, provides a solid explanation \citep{DC93, Fan93, CM1995}. By systematically observing the tilt of a large number of BMRs (over two solar cycles), starting from their early emergence on the solar surface, \citet{Anu23, sreedevi24, sreedevi25} showed that BMRs display tilt according to Joy's law from the time of their emergence on the solar surface, as seen in the magnetograms, supporting the theory of Joy's law tilt; however see, \citet{Schunker20} for an alternative views. 

(iv) Surface flux transport (SFT) models \citep{Wang1989, WNS89} and dynamo simulations \citep{KarakIAU24}, complemented by observations \citep{Ca13, Mord20, Mord22}, demonstrate that due to their finite tilt, when BMRs decay and disperse, they produce a net poloidal field as proposed by \citet{Ba61} and \citet{Le64}. Observations also show that the flux, tilt angle, and the latitude position of the BMRs in a cycle primarily determine the polar flux generated at the end of that cycle \citep{KO11, Das10, Muno13, Dey25}. The meridional flow plays a crucial role in the generation and transport of the poloidal field in the way it is observed on the surface of the Sun \citep{Bau04, VK24}, and the good thing about it is that this flow exists unambiguously \citep{hanasoge22}. 

There are, of course, several processes, including the equator-ward migration of the sunspot band, the formation of sunspots from the toroidal field, whose mechanisms are less constrained through observations; in this review, we refrain from discussing them \citep[however see,][]{NC02, Fa21, Kar26}, and now we move to the detailed modelling part of the solar dynamo.  

\section{
Dynamo models for the solar cycle prediction}
The magnetic field that we discussed above, namely, the observed poloidal and toroidal fields and the fields of sunspots and BMRs are of large-scale or global\footnote{In contrast to the local or small-scale dynamo, the length- and time-scales of the generated magnetic field in the large-scale dynamo is much larger than that of the driver---the convective flow.} type, that are often studied through the mean-field dynamo. 
In the models for mean-field dynamo, one is interested in obtaining the large-scale magnetic field, by modelling the small-scale magnetic and velocity fields using suitable approximations \citep[see][for an in-depth discussion of the mean-field electrodynamics]{KR80}. In this model, we solve the following form of the Induction equation,  

\begin{equation}
 \frac{\partial \vec{\avg B}}{\partial t} =  \vec{\nabla} \times \left( \vec{\overline{v}} \times \vec{ \avg B}  + 
 \alpha \vec {\avg B} -  \eta_t ( \vec{\nabla} \times \vec {\avg B}) \right).
 \label{eq:indm}
\end{equation}
Here, $\vec{\overline{v}}$ is the mean (axisymmetric) velocity field, arising from the meridional circulation, which gives the radial and latitudinal velocity components and the differential rotation, giving the longitudinal flow. The $\alpha$ is the $\alpha$ effect which comes from the non-zero helicity in the flows, and in the classical $\alpha \rm \Omega$ model, it is this term that is responsible for generating poloidal field from the toroidal one. The $\eta_t$ is the turbulent diffusivity, which is much larger than the molecular one and thus the latter is ignored in the above equation.  

Ideally, one needs to solve a dynamical equation for the mean/large-scale velocity $\vec{\overline{v}}$, which includes the back reactions of the large-scale and the small-scale magnetic fields \citep[e.g., Eq. 4 of][]{Kar23}. However, this equation again involves the parameterisation of the small-scale turbulence and magnetic fluctuations (e.g., through the $\Lambda$ effect and turbulent viscosity) and requires turbulence model \citep[see, e.g.,][]{Kipen63, Rudiger89, KR93, KO11b}. Fortunately, we have some knowledge of the $\vec{\overline{v}}$ from the observations. The differential rotation has been measured on the solar surface for more than a century and in the entire convection zone for several decades, and these observations also show that it varies only a little (in the form of torsional oscillation, which is less than 0.5\%) \citep{GH84, How09, Jha21}. The meridional component of the flow is known on the solar surface through observations, while its profile in the deeper convection zone is guided through MHD modelling and global convection simulations \citep{Kit16, Chou21, Kar15, KKC14, FM15, KMB18b}, supported by limited observations \citep{Gizon20}. Therefore, without solving the equation for $\vec{\overline{v}}$, one can specify its value through the observed (and guided by observations and modelling) meridional and differential rotation, which is the essence of the kinematic dynamo model.

\subsection{Kinematic axisymmetric  models}
In a kinematic dynamo, we specify the large-scale flow and solve the equation of the large-scale magnetic field to study the dynamo mechanism. The kinematic approach is not a bad prescription for the Sun, as recent evidence suggests that the solar dynamo is not highly supercritical \citep{Met16, Met25, KKB15, KN17, ghosh24, wavhal}, and thus the nonlinearity is not too strong to significantly influence the flow. Furthermore, the large-scale poloidal magnetic field, the solar corona during minima, and the large-scale flows are all observed to be predominantly axisymmetric \citep{Cha20, Chou21}, motivating modellers to consider an axisymmetric approximation in which the fields are written in the following forms. 
\begin{equation}
\vec{\overline{B}} = \vec{ B_{\rm p}} + \vec{B_{\rm t}} =  \del \times \left[A(r,\theta,t)\vec{ \hat{\phi}} \right] + B (r,\theta,t) \vec{ \hat{\phi}},
\end{equation}
where $A$ is the potential for the poloidal field: $\vec{B_{\rm p}}$, which includes $B_r (r,\theta,t) $ and $B_\theta (r,\theta,t)$ components of the magnetic field, and $B (r,\theta,t)$ is the toroidal component.
On the same footing, the velocity field can be written as
\begin{equation}
\vec{\overline{v}}(r,\theta) = \vec{v_{\rm m}} (r,\theta) +  v_\phi (r,\theta)  \vec{ \hat{\phi}}  =  v_r (r,\theta)  \vec{ \hat{r}} + v_\theta (r,\theta)  \vec{ \hat{\theta}}  + r \sin\theta {\rm \Omega} (r,\theta) \vec{ \hat{\phi}},
\end{equation}
where $ \vec{v_{\rm m}} =  v_r (r,\theta)  \vec{ \hat{r}} + v_\theta (r,\theta)  \vec{ \hat{\theta}} $ is the meridional circulation and ${\rm \Omega} (r,\theta)$ is the angular frequency.
With these forms of magnetic and velocity fields, one can derive the following equations for the poloidal and toroidal components.
\begin{equation}
\frac{\partial A}{\partial t} + \frac{1}{s}(\vec {v_m} \cdot \del)(s A)   = \eta_t\left(\nabla^2 - \frac{1}{s^2}\right)A + \alpha B,
\label{eq:pol}
\end{equation}
\begin{equation}
\frac{\partial B}{\partial t} + \frac{1}{r}\left[\frac{\partial (r v_r B)}{\partial r}+ \frac{\partial (v_\theta B)}{\partial \theta}  \right] = \eta_t\left(\nabla^2 - \frac{1}{s^2}\right)B 
+ s(\vec {B_p} \cdot \del){\rm \Omega} + \frac{1}{r}\frac{d\eta _t}{dr}\frac{\partial (rB)}{\partial r},
\label{eq:tor}
\end{equation}
where $s= r\sin{\theta}$ and $\eta_t$ is assumed to depend only on $r$.

In the above \Eq{eq:pol}, the source for the poloidal field is the $\alpha$ effect, while in \Eq{eq:tor}, it is the non-uniform rotation (second term on the RHS)---the $\rm \Omega $ effect. We note that the above equations formally describe the $\alpha \rm \Omega$ model for the Sun (we have neglected a term $\vec{\hat{\phi}} \cdot [\del \times (\alpha \vec{B_p})]$ in the derivation of above equations, which causes toroidal field generation from the poloidal field through the $\alpha$ effect).


\subsection{\bl\ and flux transport dynamo models}
As discussed in \Sec{sec:physicalmodel}, the poloidal field in the Sun is generated through the decay of tilted active regions and BMRs. The dynamo models that consider this process for the generation of poloidal field are popularly known as the \bl\ dynamo models. Historically, this \bl\ process, in the axisymmetric models, was captured by adding a term $\alpha$ which is non-zero near the surface and in latitudes where active regions are observed. Essentially, by replacing $\alpha B$ term in \Eq{eq:pol} with 
$  \alpha_{BL} (r,\theta) B_{BCZ} (\theta),
$ 
(where $B_{BCZ}$ is the toroidal field at or around the base of the CZ) various models were developed \citep{CSD95, DC99, CNC04, GDG09, KP13}; also see reviews of \citet{Kar14a} and \citet{Hazra23}.
In recent years, the \bl\ process has been captured more realistically by placing toroidal flux tubes from the base of CZ through flow perturbations \citep{YM13, Kumar19} or by depositing BMRs explicitly on the solar surface based on the toroidal field at the base of the CZ \citep{MD14, MT16, HM18}.   

Usually these \bl\ dynamo models include a meridional circulation, which is pole-ward near the surface (with a speed of about 20 \mps\ peaking near mid latitude) and equator-ward at the base of the CZ of speed about a few \mps\ \citep{WSN91, HKC14}. This meridional flow advects the toroidal field at the base of the CZ from high to low latitude, which causes equator-ward migration in the toroidal field (or proxy of sunspots) in these models. The \mc\ also helps generate a poloidal field by efficiently transporting the leading polarity fluxes of BMRs towards the poles \citep{Wang1989}.  We emphasise that although \mc\ plays an important role in \bl\ dynamo models, and thus they are also named flux transport dynamo models, any dynamo model in which the \mc\ is the primary mechanism of the flux transport is called the flux transport dynamo model \citep{Cha20}.  

\section{Surface Flux Transport (SFT) model for prediction}
\begin{figure}
    \centering
     \includegraphics[width=0.75\linewidth]{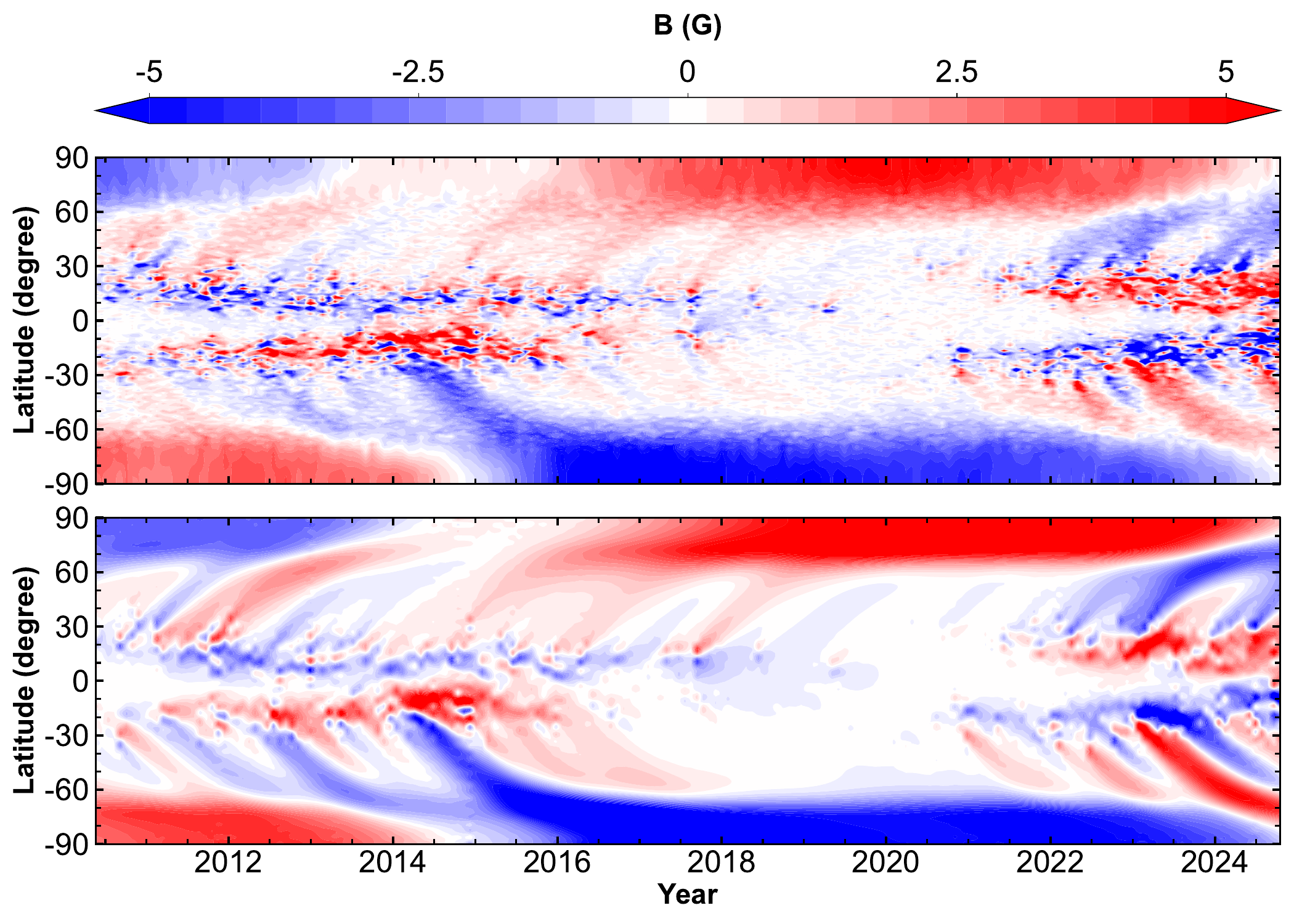}
    \caption{The surface radial magnetic field obtained from a SFT model of \citet{Ruihui25} (bottom) and the observed field from the Helioseismic
and Magnetic Imager on board the Solar Dynamics Observatory (HMI/SDO; bottom). The figure reproduced with permissions from authors. 
    }
    \label{fig:sft}
\end{figure}

The magnetic field on the solar surface is observed to be dominated by the radial component \citep{WS92}, which is an important component of the poloidal field and acts as the seed for the next sunspot cycle. Hence, the spatial and temporal evolution of the radial magnetic field on the solar surface helps to understand the generation of the poloidal field --- the  \bl\ process. In the SFT model, the radial magnetic field that acts as a passive scalar field is evolved through the following equation \citep{yeates23}.
\begin{equation}
    \frac{\partial B_r}{\partial t} + \bm \nabla \cdot (\bm u_h B_r)  = \eta_t \nabla^2 B_r + S(\theta,\phi,t),
\label{eq:sft}
\end{equation}
where, $\bm u_h$ is the steady (axisymmetric) horizontal surface flow that includes the meridional flow ($v_\theta (R_\odot,\theta)  \vec{ \hat{\theta}} $) and the differential rotation ($R_\odot \sin\theta \ {\rm \Omega} (R_\odot,\theta) \vec{ \hat{\phi}}$). The $\eta_t$ is the turbulent diffusivity, and  $S(\theta,\phi,t)$ is the term that represents the deposition of BMRs. As the plasma flow, diffusivity and the magnetic field of BMRs are all reasonably measured on the solar surface, the SFT model has proven success in reproducing the evolution of the surface magnetic field (\Fig{fig:sft}). In the past, various SFT models have been used to reconstruct the historical magnetic field utilising the sunspot group observations \citep{jiang11, yeates25, jha25, gopal25} and for the prediction of the solar cycle \citep{Iijima17, jiang2018predictability, UH18, BN18, BKK23}, which we discuss in the next section. We note that the 3D dynamo model, STABLE, also captures the essence of the SFT model, as this dynamo model includes BMRs as the source for the poloidal field and realistically models the surface radial field \citep{MT16, HCM17}.

\section{Physics of solar cycle prediction}
Already from the observed variation of the polar magnetic field and the sunspot number, as shown in \Fig{fig:rad2ssn}, we find a close correspondence between the strength of the polar field, which becomes strong at the solar minima, and the amplitude of the next cycle, which peaks after about 6 years. Further, the peaks of the polar field at/around solar minima show a strong correlation with the amplitude of the next solar cycle.  This causal connection is also supported by the \bl\ dynamo model discussed above; see also \Eq{eq:pol}. In this scenario, the polar flux is wound by a steady differential rotation inside the convection zone, which emerges as the sunspots for the next cycle (\Sec{sec:physicalmodel} for more details).  We note that the strong correlation/dependency of the toroidal field on the previous polar field is insensitive to the details of the dynamo model \citep{CB11, kumar21b, Li26}.  Any dynamo model, including the non-\bl\ type models, has this predictive capability using the poloidal field. However, the delay between the poloidal and sunspot is determined by the flux transport process, and thus depends on the model parameters \citep{JCC07, YNM08, KN12}. In \bl\ models, the transport of the field is largely due to the meridional flow (unless the diffusivity is too strong), which sets the correct time delay of about 5 years between the polar field and the peak of the toroidal field \citep{YNM08}. The toroidal to sunspot process is anyway short \citep{Jouve10}, as the buoyant rise of the toroidal flux tubes occurs on a time scale of months and does not contribute to the poloidal to toroidal timescale. 

\section{Predictions utilizing the polar field at/around the solar minimum} 
\begin{figure}
    \centering
    \includegraphics[width=0.85\linewidth]{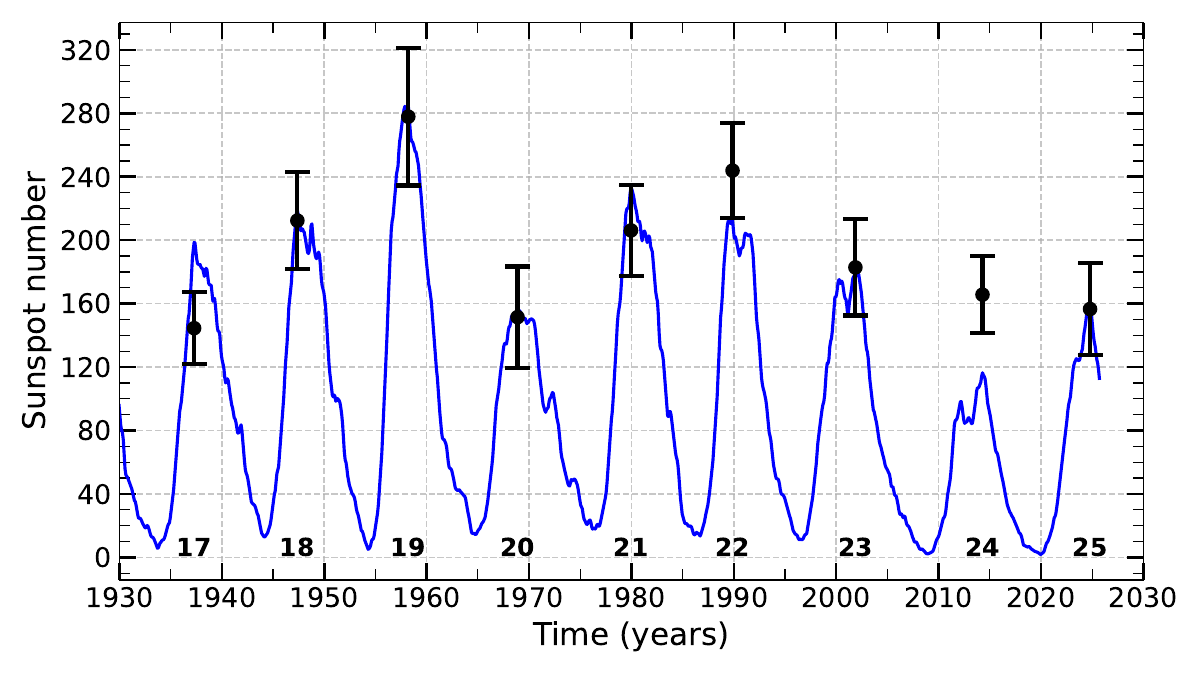}
    \caption{Predicted peaks of the sunspot cycles, marked by the dots, using the observed correlation between the polar field/proxy and the peak sunspot number. Error ranges are taken as one $\sigma$ deviation of the linear regression. Polar field data is homogenised using the WSO polar field measurements for Cycles 23--24 and A(t) index for Cycles 15--22 \citep{Makarov01}. 
    }
    \label{fig:predic_pol}
\end{figure}

The physics-inspired strong correlation between the observed polar field (or its proxy) at the solar minima and the amplitude of the next cycle allowed researchers to predict the amplitude of the solar cycle. The first prediction in this method was made by \cite{Sch78} for Cycle 21, followed by the predictions of \citet{Schatten2005}, \citet{Svalgaard2005}, \citet{JCC07}, and \citet{Pishkalo2010} for Cycle 24, and \citet{Pesnell2018}, \citet{Svalgaard20},  \citet{Janssens21} and \citet{Kumar21} for Cycle 25 (more in \Sec{sec:conclusion}). All these predictions were not too far from reality. The deviation in the predicted values is primarily due to the use of different polar field (or proxy) data to generate the linear relation and the time at which the polar field/proxy is measured. As this linear relation is not perfect and there is a considerable amount of scatter around it, the uncertainty in the predicted sunspot number is also considerable.  To demonstrate how the predicted values corroborate the reality, we show the predicted values of the peaks of the past Cycles 15--22 in \Fig{fig:predic_pol}, obtained using the linear relation between the polar field at the solar minima and the amplitude of the next sunspot cycle. The large mismatch in some of the cycles suggests that one needs to carefully estimate the polar field at the solar minimum or there is additional (missing) physics involved in this process.   

\section{Predictions from the dynamo models utilising the polar field} 
Instead of utilizing the observed linear correlation between the polar field at the minimum and the amplitude  of the next cycle,
one can employ a dynamo model to predict the sunspot number (or its proxy from the toroidal field) for the next cycle from the observed polar field. This is possible because in the solar dynamo model, as detailed in \Sec{sec:physicalmodel}, the polar flux to the toroidal flux (which gives the sunspot for the next cycle) is predictable. Kinematic dynamo models of \bl\ type were indeed used to predict the strength of the solar cycle by utilising the observed polar field at the solar minimum. The first prediction of this type was made by \citet{CCJ07}, who fed the observed dipole moment of Cycles 20--24, computed from the Wilcox Solar Observatory (WSO) and Mount Wilson Observatory (MWO), into their flux transport dynamo model to model Cycles 21--23 and predict Cycle 24. Later, \citet{JCC07} used the radial field from WSO measurements to improve the prediction. For the current solar cycle, a similar approach was made by \citet{Guo21}. These predictions are closer to those made simply using the linear correlation between the polar field and the amplitude of the next sunspot cycle. 

We note that the predicted values from these two methods, although based on the same physics, may not necessarily be the same. There is a weak memory of the polar field that can be propagated from one cycle to the next (detailed presentation of this in \Sec{sec:multicycpred}.) It is not easy to answer whether the use of a dynamo model improves the prediction of the polar field. However, as the Sun's polar field has a weak memory, and if it is captured correctly in a dynamo model, then the dynamo model can undoubtedly do a better job. Dynamo-based prediction also has the advantage of capturing the variation of the \mc, which is observed on the solar surface. Research shows that a change in the deep \mc\ can lead to a significant variation in the strength of the predicted toroidal flux \citep{YNM08, Kar10, KC11, VK24}, emphasising the need to capture the variation of the deep \mc\ in prediction models. \citet{HC19} captured this variation by embedding the decay rate of the previous cycle \citep[which they argued is due to the variation of \mc;][]{Haz15}, into the polar field and predicted the amplitude of solar cycle 25. As their predicted value is closer to reality compared to the sole polar field-based prediction, we tend to accept the need for capturing the variation of \mc\ in the dynamo model for better prediction of the solar cycle.          Another advantage of dynamo-based prediction is that it allows us to predict the full profile of the predicted sunspot cycle. In contrast, the polar field correlation-based prediction only predicts the amplitude of the next cycle, and additional information--such as the Waldmeier law--is required to construct the full profile.

\section{What limits the accuracy of the polar field-based predictions?}
Now we discuss why we cannot accurately predict the amplitude of the cycle using the polar field at/around the minimum.
The first obvious answer to this question is the uncertainty in measuring the polar field. Due to the line-of-sight effect, any measurements of the polar regions are highly compromised. Even if we have reliable measurements of the magnetic field in the polar regions, we have uncertainties in defining the time of the polar field for the prediction, whether it is the solar minimum or at an earlier or later time. Studies show that the polar field indeed starts to exhibit a strong correlation a few years before the solar minimum \citep{Kumar21}, but these studies are based on limited cycles, and it is not possible to uniquely determine the exact time of the polar field to be used for prediction. 

The next issue is the limited number of cycles (last four cycles only) for which the polar field measurements are available. From the proxy, we have somewhat longer duration for the polar field, namely the A(t) index of \citet{Makarov01} is available from cycle 15 to 22, the number of polar faculae \citep{Munoz2012} and the polar active networks \citep{Priy14, mishra25} are available from cycle 15 to 22. However, all these proxy data are not very accurate and irregular, which causes considerable scatter in the linear relation between the polar field (proxy) at the minimum and the amplitude of the next cycle, resulting in uncertainty in the prediction (\Fig{fig:predic_pol}).    

When it comes to dynamo-based prediction, there are also issues. The first one is the implementation of observed data into the model. The past methods \citep[e.g., correcting the poloidal field above $0.8 R_\odot$ using the observed surface value at the solar minimum as done in][]{CCJ07} are not satisfactory; however, see \citet{gopal25}, for a promising method of feeding the observed magnetic field into the 3D dynamo model that can be extended for the cycle prediction.
The second is the not-so-well-constrained parameters of the model (more discussion on this is in \Sec{sec:multicycpred}). 
The third is the assumption of the linearity between the polar field and the next cycle strength. Although the observed data of polar field and its proxy show a `reasonable' linear trend, a weak nonlinearity and stochastic components cannot be ignored \citep{BKC22}. This departure from a perfect linear trend could be due to considerable variation in shear, as found in helioseismic observations \citep{ABC08}. 
Finally, we have the variation in the meridional flow, of which the surface observations indeed show some variation \citep{Gonz06, Gonz10, GR08, BA10, HR10}.
In the \bl\ dynamo model, in addition to transporting the fields, \mc\ also affects the strength of the field. Strong flow can also drag the leading polarity flux to the poles, reducing the efficiency of polar field generation \citep{VK24}. Furthermore, strong meridional flow transports the magnetic field fast in the deep CZ, allowing the diffusion to reduce its strength \citep{YNM08, Kar10}. Thus, any temporal variation in \mf\ can potentially limit the predictability of the polar field \citep{UH14b, HC19}. 

\section{Predictions before the solar minimum: how early can we go?}
As discussed above, the usual `landmark' of the solar cycle prediction is the preceding solar minimum, when the polar field has reached its maximum value or has stopped growing. However, predicting the solar cycle even before the solar minimum is always demanding, and for this, we need to estimate the polar field strength before the end of the cycle. It is the decay of the BMRs that produces the polar field for the next cycle. The BMRs from about the first half of a cycle are used to cancel the existing polar field (of the previous cycle), and the remaining BMRs are spent to generate the polar field for the next cycle, which we need for the prediction. Hence, we have two options: the first one is to wait for a few more years after the solar maximum (i.e., to pass the solar maximum epoch unambiguously) to entirely cancel the previous polar field and observe how the polar field for the next cycle is growing. The rate of growth of the polar field can indicate the next cycle. Another option is to replicate the remaining half of the cycle after it has started decaying, using an identical cycle from the past, or simply fill up the remaining half using a synthetic cycle.   Both of these options were taken by the prediction community and we elaborate on them below. 

\begin{figure}
    \centering
    \includegraphics[width=0.75\linewidth]{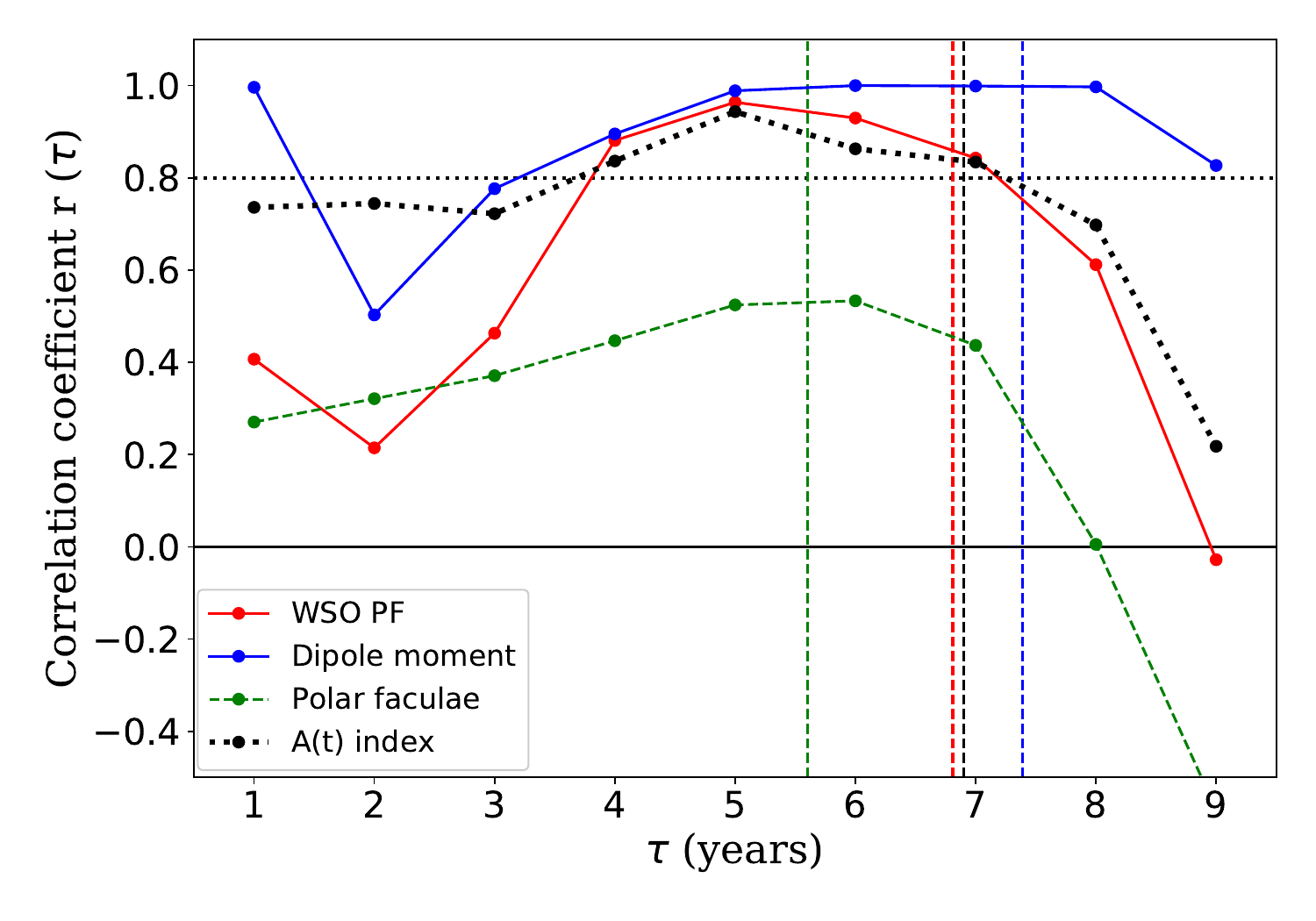}
    \caption{
Linear (Pearson's) correlation coefficient between the polar field (or proxy) at different times $\rm \tau$ measured forward from the time of reversal and the peak sunspot area of the next cycle.  Note $\tau = 0$ corresponds to the time of polar field reversal. Vertical dashed lines of different colour mark the time shift from the reversal of the polar field (or proxy) to the cycle minimum. \citet{Kumar21} for more details. 
}
    \label{fig:pf_tau}
\end{figure}

\subsection{Prediction using the growth rate of the polar field}

\begin{figure}
    \centering
    \includegraphics[width=1.0\linewidth]{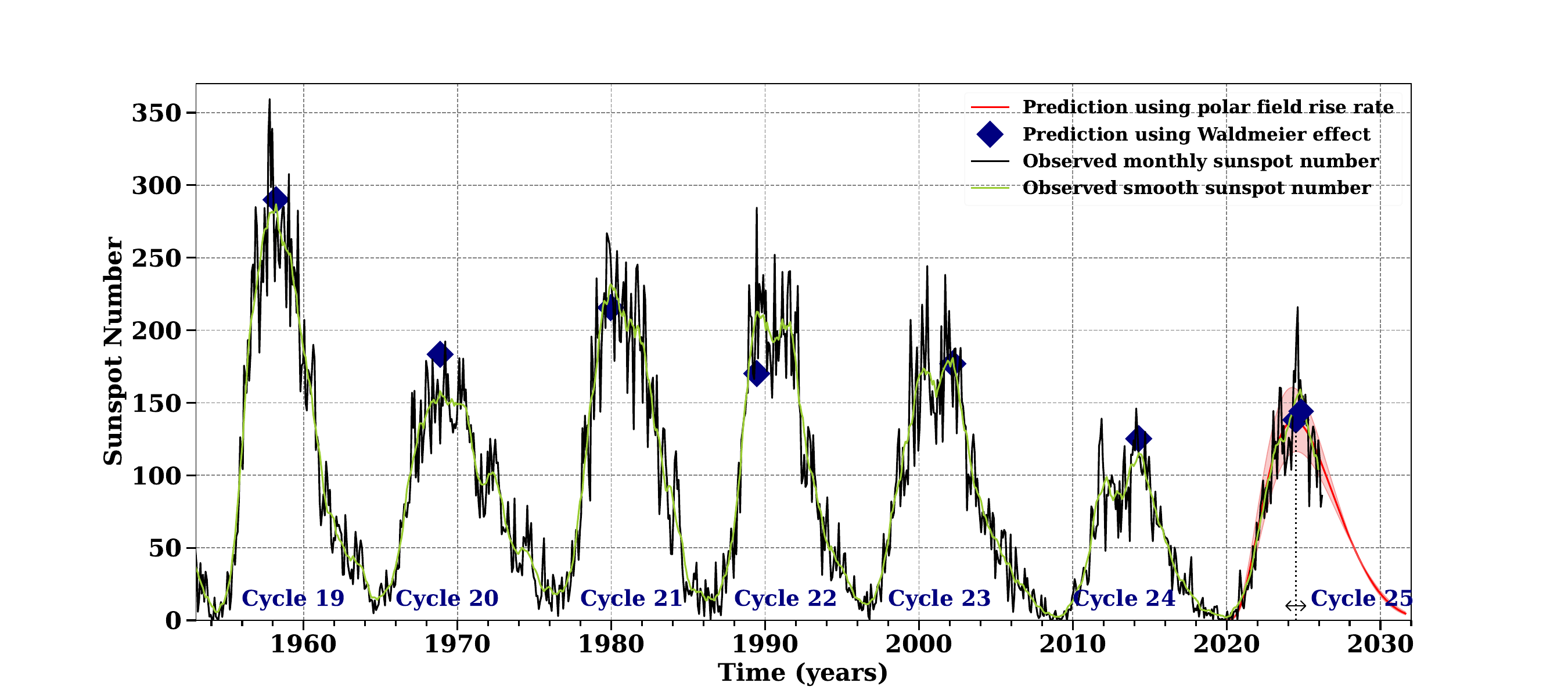}
    \caption{The predicted value of the Cycle~25 (red curve) using the rise rate of the previous polar field (till 2016) as proposed in \citet{Kumar22} and its comparison with the observed sunspot data (black and green curves, updated on March 2026). The predictions based on the rise rate of the sunspot cycle (the Waldmeier effect) are shown with filled squares. 
}
    \label{fig:ourprediction}
\end{figure}

After the solar maximum, if we wait for another 2--3 years, then one can get an indication of the strength of the solar cycle by measuring how rapidly the polar field is growing. \citet{Kumar21} showed that we do not have to wait until the solar minimum to make a prediction of the next cycle; even after about 4 years of the reversal, the polar field begins to correlate with the amplitude of the next cycle, as shown in \Fig{fig:pf_tau}. This means that one can even make a prediction about 2-3 years before the solar minimum, the usual landmark when the polar field reaches to its peak.  Later, after careful analysis of the rise rate of the polar field, computed from two-year data after one year of reversal, \citet{Kumar22} demonstrated that the rise rate reasonably determines the amplitude of the polar field and thus the strength of the next cycle. They had shown the robustness of their results using the \bl\ type dynamo model and the SFT model with synthetic solar cycles and observed variations in BMR properties \citep{BKK23}. They further showed that the prediction based on the Waldmeier effect --- strong cycles rise faster than the weaker ones \citep{wald, CS08} --- coincides with the prediction based on the rise rate of the polar field. It is really exciting to see that their prediction is not too far from the reality as presented in \Fig{fig:ourprediction}; also see Figure 2 of \citet{Kumar22} for details. Although encouraging, the polar field growth-based early prediction of the solar cycle cannot be as accurate as the prediction based on the polar field at the solar minimum, due to the irregularities involved in the polar field generation process, and there may be some ambiguity in determining the time of reversal for some cycles \citep{Mord22, Golubeva23, Jha24}; more on this is in Sec~12.

\subsection{Prediction using SFT model}
As shown by \cite{wald, Hat15, CS16, Kar26}, the declining phase of the solar cycle is largely cycle-independent. Hence, once the activity level starts to fall, the remaining portion of a cycle can be reasonably reconstructed using the properties of past cycles. \citet{Hat15} and \citet{jiang2018predictability} have given a detailed procedure of how to generate a (synthetic) profile of the solar cycle. They further showed that the prediction of the sunspot group emergence is possible after a cycle has passed about four years into that  cycle. Finally, they applied their method for predicting the emergence of sunspot groups for the last two years of Cycle 24, and fed them into their SFT model to predict the amplitude of Cycle 25 by relying on the correlation between the axial dipole moment and the amplitude of the next cycle.

Using a similar method, \citet{BN18} independently generated the BMRs with several realizations and observationally-guided variations in their properties for the last 3.25 years of Cycle 24 (the remaining phase of the cycle when they made the prediction) and fed them into their SFT model to predict the polar field at the end of the cycle. Subsequently, they had assimilated the magnetic field from their SFT model into an axisymmetric dynamo model, through which they predicted the toroidal flux for the next cycle. 
\citet{Labonville2019} took a similar path but they used a data-driven coupled $2 \times 2$ D flux transport dynamo model to predict the amplitude of Cycle 25 a few years before the solar minimum. 

Fundamentally, the prediction methods of \citet{jiang2018predictability} and \citet{BN18} are, by and large, similar, and their predictions are close to each other. A slightly better match of the former model could be due to different parameters in their SFT models and the different process of translating the polar field into the next cycle amplitude. However, a considerably lower value of the predicted sunspots for Cycle 25 in \citet{Labonville2019} could be due to optimisation of the parameters, which was based on the observed data for cycles 23-24.

\begin{figure}
    \centering
    \includegraphics[width=0.85\linewidth]{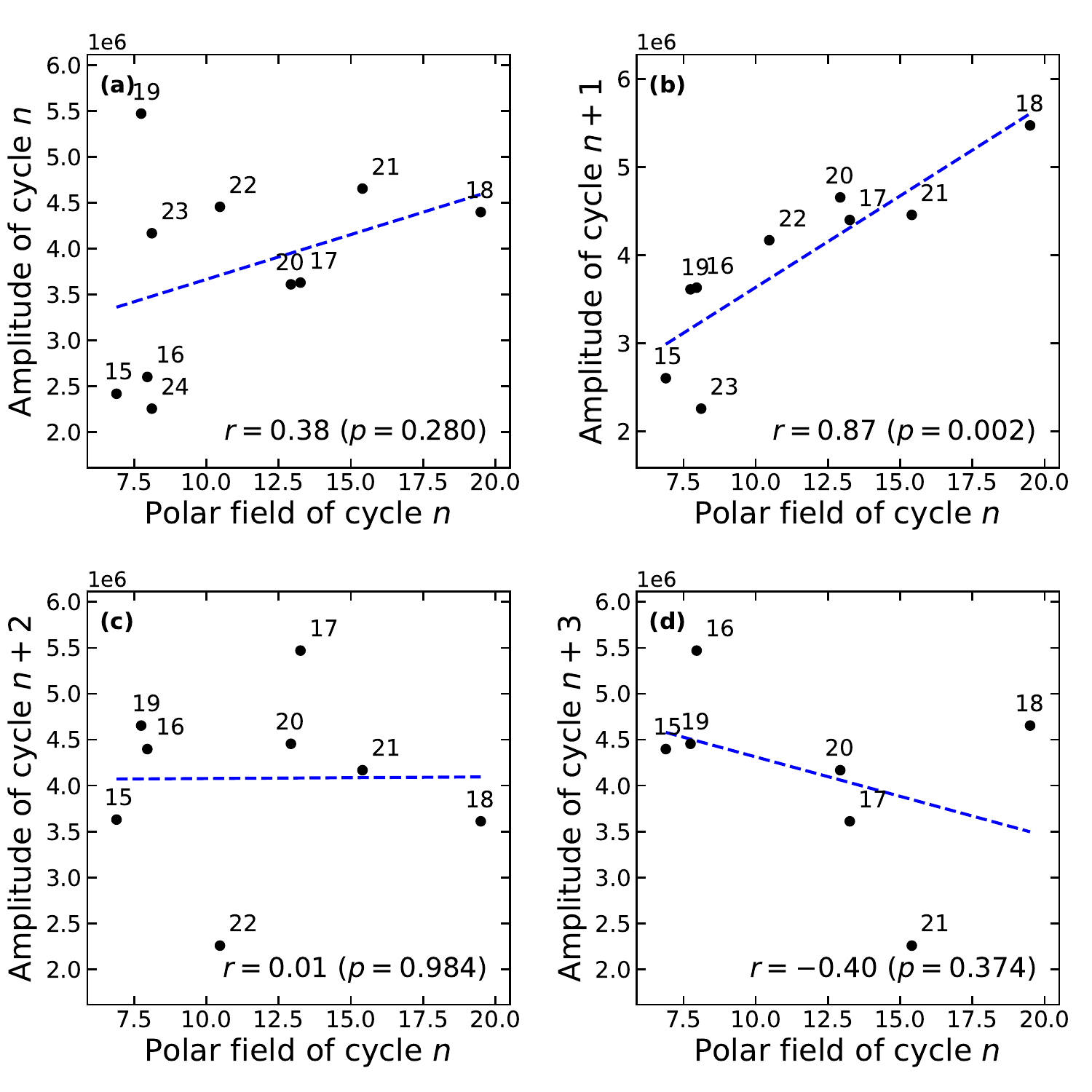}
    \caption{
    The scatter plot between the peak of the polar field at cycle $n$ versus the amplitude of cycle $n$ (a), $n+1$ (b), $n+2$ (c), and $n+2$ (d). The cycle number ($n$) corresponding to each point is labelled next to it. The cycle amplitude is obtained from the calibrated group area of \citet{Mandal2020}.  Here, the polar field is obtained from the A(t) index of \cite{Makarov01} for cycles 15--22 and WSO for cycles 23--24, scaled appropriately to bring it to the A(t) value. 
    }
    \label{fig:pol2ssn_multi}
\end{figure}

\section{What limits the prediction beyond one cycle}
\label{sec:multicycpred}
From the discussion in the previous section, we learned that only after a few years of the solar maximum we can reasonably predict the amplitude of the next cycle by reconstructing the BMR properties for the remaining cycle or simply by measuring the growth rate of the polar field. The more time we spend in a cycle, the less uncertainty there will be in the polar field (due to less uncertainty in the BMR properties).  Now the question is, if we know the polar field of a cycle, can it not predict more than one cycle? The answer to this question may be given by analysing the correlation between the polar field at the solar minimum and the amplitude of the subsequent cycles. Unfortunately, the observed polar field is available only for the last four cycles; thus, the multi-cycle correlation is not reliable solely based on the polar field. \citet{Muno13} analysed the facula count---a measure of the polar field---for Cycles 14--22 and found insignificant correlation of the polar field of cycle $n$ with the amplitude of cycles $n+2$ and $n+3$; also see \citet{Nandy21}.  We perform an independent check utilising a composite data produced from the A(t) index of \citet{Makarov01} and WSO polar field, and the results are shown in \Fig{fig:pol2ssn_multi}. It again shows that the polar field has no significant correlation beyond one cycle. 

\begin{figure}
    \centering
    \includegraphics[width=1.0\linewidth]{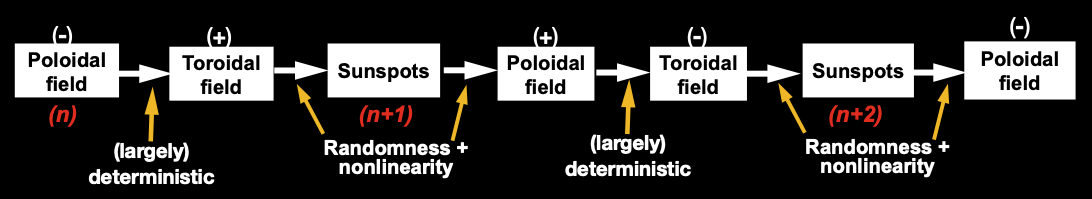}
    \caption{The basic dynamo chain. The poloidal field of cycle $n$, assumed to be negative (inward) in the north pole, generates a toroidal field of east-west direction in the northern hemisphere (through the $\rm \Omega$ effect), which erupts as sunspots for the next cycle $n+1$. The sunspots decay to generate a poloidal field of opposite polarity (with respect to the existing polarity) through the so-called \bl\ process. The cycle then continues. 
    We note that as long as the meridional flow is assumed to be steady, the poloidal to toroidal conversion is deterministic, since the differential rotation shows only a tiny variation. However, the toroidal field to sunspots and their subsequent decay to the poloidal field involve some stochasticity and nonlinearity. 
    }
    \label{fig:dynamo_loop}
\end{figure}

To understand the physics of why the correlation is limited to only one cycle, we recall the dynamo mechanism as illustrated in \Fig{fig:dynamo_loop}. We begin with a poloidal field of a cycle ($n$), the winding of which through the differential rotation generates a toroidal field. This process is deterministic and forms the basis for solar cycle prediction. However, the toroidal to sunspots for the next cycle ($n+1$) and their subsequent decay to the generation of poloidal field involve some randomness and nonlinearities, which we elaborate below.

\subsection{Fluctuations in the toroidal to poloidal field generation}
While it is known that the toroidal field gives rise to active regions/BMRs, their individual properties cannot be predicted \citep{Fa09, Cheung14}. For example, the positions of active regions on the solar surface vary significantly, although they are clustered in a narrow band of mean latitude less than about $25^{\circ}$ \citep{CS16, BKC22}. If an active region appears at high latitude, the cross-equatorial cancellation of the leading polarity flux is less, and both polarity fluxes are transported towards the pole, resulting in a weak polar flux \citep{Pet20}. SFT simulations \citep{JCS14, J20, yeates23} and dynamo model \citep{KM18, KKS24} have demonstrated that the generated polar flux and the axial dipole moment exponentially decrease following a Gaussian profile with an increase of the latitude of emergence of the active region. Next, the tilt of the active regions causes a considerable variation in the polar field \citep{KM17, Nagy17}. Observations show that a significant fraction of BMRs (about 10\% and 30\%, respectively) are of anti-Hale Joy (leading polarity is of opposite polarity and appears near the equator) and the Hale anti-Joy (the leading polarity near the pole and of correct polarity) \citep{SK12, MNL14, sreedevi26}. These `wrongly tilted' BMRs produce opposite polarity field (with respect to the dominant/expected field) in a cycle, thereby producing fluctuations in the polar field. Numerous studies using dynamo and SFT models have demonstrated the variation in the polar field due to this tilt anomaly and its implications in modelling the solar cycle irregularity \citep{Kar23}. The analysis of the observed surface magnetic field and the active region tilts also reveals fluctuations in the polar field caused by the wrong orientations of the active regions \citep{Ca13, Mord20, Mord22}.  The next is the flux of the active region, which is also not a fixed quantity, and its value is given by a distribution as found in the observations. Finally, the varying properties of the time delay between the emergence of two subsequent active regions and the separation between two poles of an active region also cause disturbance in the generated polar field \citep{KKS24, Kumar26, sreedevi26}.    

\subsection{Nonlinearities in the toroidal to poloidal  field generation}
The next is the nonlinearity in the generation of the poloidal field. If a cycle is strong, its toroidal field in the deep CZ is high, and it generates a higher number of sunspots. Then, we vaguely expect the polar field generated at the end of this cycle to be high. Therefore, a strong cycle should follow another strong cycle, assuming that the effects of fluctuations in the polar field generation process is not too strong. However, this linear trend from one cycle to the next will not hold for a very strong cycle. Observations display that a strong cycle starts generating sunspots at higher latitudes with respect to an average cycle \citep{SWS08, MKB17}. And, from theoretical modellings \citep{JCS14, KKS24}, we know that high-latitude sunspots are less efficient in creating the polar field. Hence, if a cycle is strong, which generates a higher number of sunspots, we get a weaker polar field at the end of the cycle. Thus, the generated polar field at the end of a cycle should depend `nonlinearly' on the cycle strength. This nonlinearity, named `latitude quenching' \citep{Petrovay20, J20}, is a potential candidate for regulating the polar field in the Sun \citep{Kar20, Talafha25}.  \citet{BKC22, Kar26} showed that the latitudinal quenching is caused by the nonlinear flux loss from the interior due to magnetic buoyancy. A strong cycle releases flux at a faster rate in its early phase when the activity belt is already in high latitude. 

Another potential candidate that can also introduce nonlinearity in the polar field is the tilt quenching. The thin flux tube model \citep{s81, LK97} suggests that a strong toroidal flux tubes rise fast through the convection zone, allowing the Coriolis force to act on them for a lesser time \citep{DC93, FFM94}. Thus, an intense high-flux sunspot should have less tilt, which is generally known as the tilt quenching \citep{LC17, KM17}. Due to measurement challenges in determining the tilt from observed magnetograms and a significant amount of scatter associated with the measured tilt angle, the tilt quenching is not well-constrained observationally \citep{Jha20, sreedevi24, qin25}. However, when we measure the cycle-average tilt, we clearly observe a negative trend with the cycle amplitude \citep{Das10, Das13, Jiao21}. 
In summary, similar to latitude quenching, the tilt quenching again tends to generate a weak cycle following a strong cycle, which was demonstrated in SFT simulations by \citet{J20} and \citet{Talafha22} and in observations by \citet{Dey25}.

Other nonlinear processes may be involved in generating the poloidal field. These at least include nonlinear inflow around active regions \citep[a strong active region produces strong inflow around it, which inhibits the cross-equatorial cancellation of the leading polarity flux;][]{CS10, kinfe24,Talafha25b}, meridional flow perturbation caused by Lorentz force \citep{KC12} or inflow \citep[a strong cycle tends to produce a large reduction of flow;][]{Jiang10}. All of these processes, by and large, reduce the polar field from the expected value when a cycle is sufficiently strong. This was demonstrated in observation as well through careful analysis of the sunspot and polar field data in \citet{Dey25}.

\begin{figure}
    \centering    
    \includegraphics[width=0.8\linewidth]{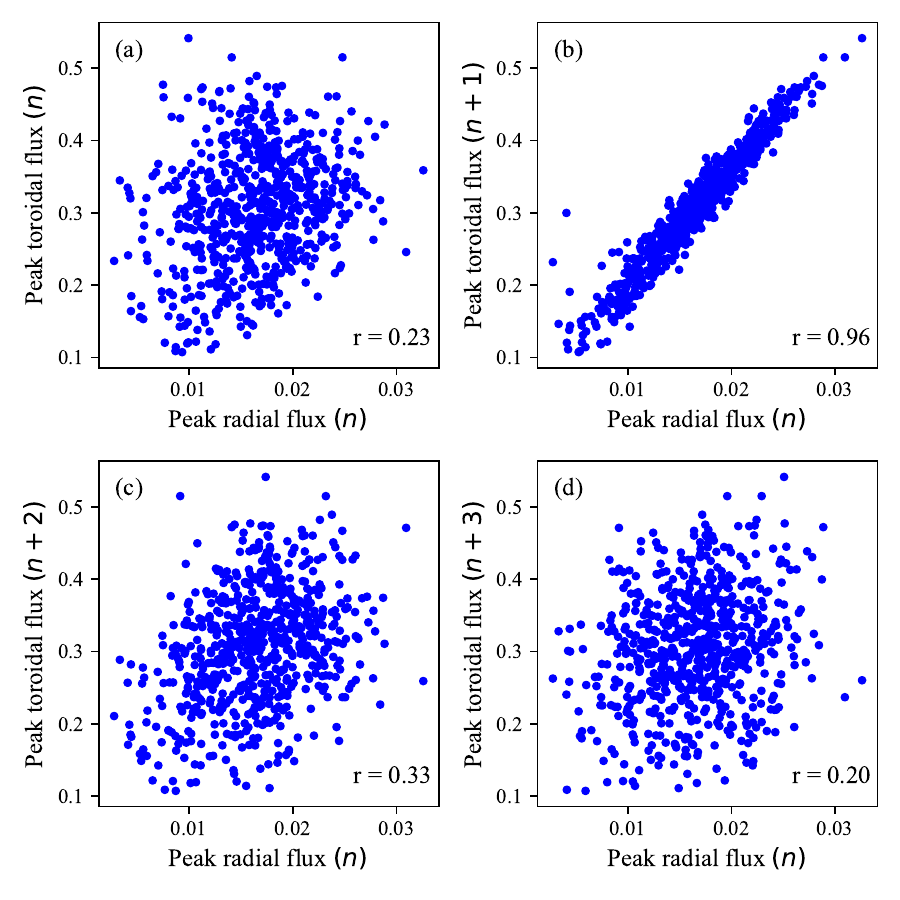}
    \caption{
    The scatter plots between the peak of the polar flux of a cycle $n$ versus the amplitude of the toroidal flux (a proxy of sunspot cycle) of cycle (a) $n$ (same cycle), (b) $n+1$ (next cycle), (c) $n+2$, and (d) $n+3$ for a dynamo simulations with stochastic $\alpha$ and amplitude limiting nonlinearity. This model operates at twice the critical $\alpha$ (weakly supercritical regime), which is supported by observations, as demonstrated by \citet{wavhal}.   
    }
    \label{fig:corr_supercriticality}
\end{figure}

So, combining all these stochastic and nonlinear processes involved in the generation of the toroidal to polar field generation part of the dynamo loop (\Fig{fig:dynamo_loop}), we can confirm that it is this process where the dynamo chain is disturbed. Thus, the memory of the polar field of cycle $n$ cannot be propagated fully to the polar field of cycle $n+1$, and the amplitude of sunspot cycle $n+2$ cannot be predicted reliably. For a weak and moderate cycle, the nonlinearity is weak, and the polar field will not differ significantly from the expected one; nevertheless, stochastic fluctuations in the BMR properties can still break the memory of the polar field substantially. Dynamo simulations with weak stochastic fluctuations and nonlinearity show that the memory of the polar field is limited to one cycle when the dynamo operates in a supercritical regime, where the effects of nonlinearity and stochasticity are more pronounced. However, a memory is propagated beyond one cycle when the dynamo operates in a weakly supercritical regime (\Fig{fig:corr_supercriticality}), which the observations and independent theoretical exploration favour \citep{ghosh24, wavhal}. However, this weak memory of the polar field that is propagated beyond one cycle is not useful for making a meaningful prediction of the solar cycle.

\section{Conclusions and future outlook}
\label{sec:conclusion}
An increasing number of predictions for solar cycle 25 clearly demonstrates that the community has realised the urgency of prediction and is trying hard to achieve a consensus. However, a significant deviation of the means of all predictions for both cycles 24 and 25 with respect to reality implies that we need to improve our predictive skills. This also poses a challenge to the solar cycle prediction panels, such as the NASA-NOAA Prediction Panel, on how to publish a consensus prediction \citep{Uptonpanel19, Miesch25}.  The extrapolation methods, including the nonlinear curve fitting, performed fairly well only after the cycle passes the solar minimum and spends a year in its ascending phase, which is then determined by the Waldmeier effect \citep{Kumar22}.  

\begin{figure}
    \centering
    \includegraphics[width=0.75\linewidth]{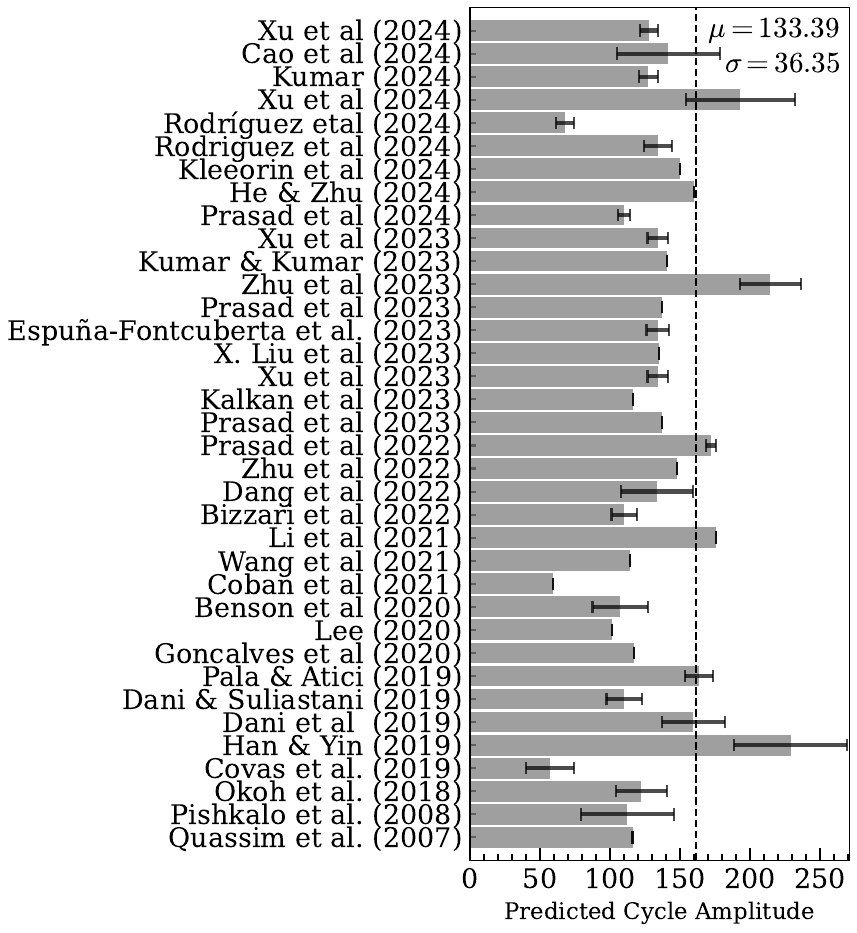}
    \caption{Predictions of the peak sunspot number for solar Cycle 25 made using ML Models. The references are in \Tab{tab:cyc25}. The $\mu$ and $\sigma$ respectively represent their mean and standard deviation. The dashed line marks the observed peak at 161. }
    \label{fig:ml_predic}
\end{figure}

In recent years, due to the development of various Machine Learning (ML) models and their increasing awareness in the solar physics community, a large number of predictions  (36 out of 136 or 26.5\%) for Cycle 24 were made employing Machine Learning (ML) models; \Fig{fig:ml_predic}. However, these predictions are also not up to the mark. The mean ($\mu$) and standard deviation ($\sigma$) of the predicted values are 133 and 36; recall that the observed value is 161. However, with respect to predictions using other methods, including those based on polar field, the ML prediction performs slightly better (recall $\mu$ = 132 and $\sigma = 39$ for all predictions for Cycle 25). However, again, the ML prediction performs better only after the solar cycle has spent its initial few years.

\begin{figure}
    \centering
    \includegraphics[width=0.75\linewidth, angle = 0]{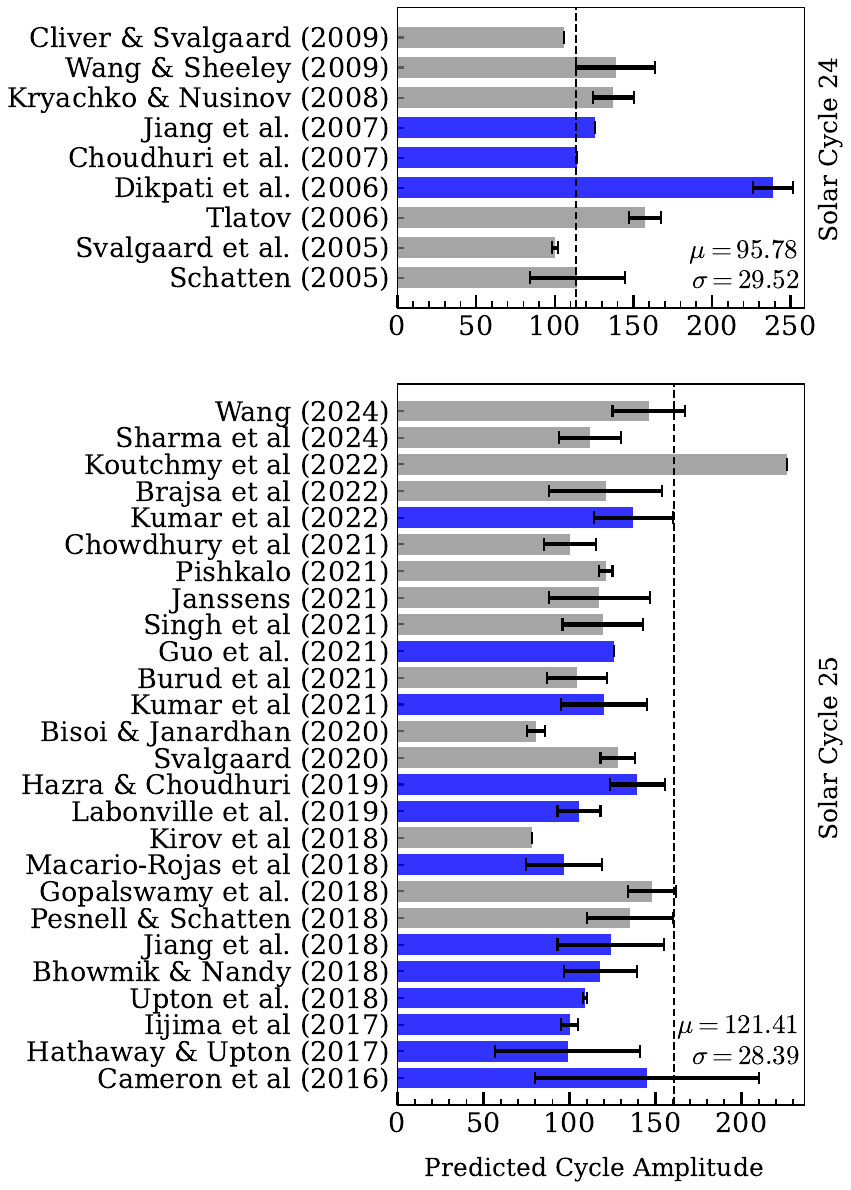}
    \caption{Predictions for Cycles 24 (top panel) and 25 (bottom) based on the polar field (or its proxy) at or around the solar minimum. Even the models that utilised a polar field or predicted/modelled the polar field (marked in blue) are included in this category because of their shared physical basis. The figure format is the same as \Fig{fig:ml_predic}. }
    \label{fig:pf_predic}
\end{figure}

The dynamo theory and various observations of the polar magnetic field and its proxy demonstrate that the polar field around the solar minimum acts as the seed for the solar cycle.  Thus, it is not unexpected that the polar field (or its proxy) based predictions perform better than other methods. In \Fig{fig:pf_predic} top panel, this is indeed what we find. This overwhelming success encouraged scientists to make predictions from multiple independent groups based on the polar field or its proxy for the Cycle 25. Nearly 19\% of predictions belong in this category. Even models that utilised the observed polar field or produced/modelled the polar field are considered in this category.   As shown in the bottom panel of  \Fig{fig:pf_predic}, the polar field-based predictions for Cycle 25, however, did not perform well.  Almost all the predictions (except those made in the last few years) did not accurately predict the peak (the mean and standard deviation are 121.9 and 27.7, respectively). 

One reason for this deviation could be the utilisation of the polar field before it has completely grown, leading to an inaccurate measure of the polar precursor. Studies demonstrate that polar field-based predictions may yield incorrect results when applied well before the solar minimum \citep{jiang2018predictability, Kumar21}. Another reason could be the real measurement problem of the polar field and its proxy (such as polar faculae). Existing measurements are highly compromised due to the projection effect; imaging the magnetic field from high latitudes will certainly help.  

Dynamo models of \bl\ type can certainly benefit by employing good quality data from dedicated polar missions and a better data assimilation technique. There may be additional physics just beyond the polar field around the solar minimum that determines the amplitude of the next cycle. The same prediction method of \cite{CCJ07} and \citet{JCC07}, developed for solar Cycle 24, performed poorly when applied to Cycle 25 \citep{Guo21},  emphasising the need to improve the dynamo model by capturing additional physics, such as variation in meridional flow \citep{HC17} and the regime of operation of the solar dynamo \citep{wavhal}.

Past predictions, complemented by dynamo simulations, suggest that making a prediction a few years before the solar minimum is possible \citep{BN18, jiang2018predictability, Kumar21, Kumar22, BKK23}; however, the appearance of anomalous active regions (such as anti-Hale and anti-Joy BMRs) increases the prediction uncertainty. Further, the synthesis of active regions into a model for the remaining period of the solar cycle introduces an additional layer of uncertainty, even after careful investigation of the solar cycle properties and the scaling relation of active region properties \citep{Nogueira26}.  Thus, any early predictions utilising the polar field before the solar minimum need to be updated with time-to-time data of the polar field for greater reliability \citep{UH18}. Both observations and the current understanding of dynamo theory suggest that making a reliable prediction before the previous cycle's maximum is not possible \citep{KN12, jiang2018predictability}. We expect that solar cycle prediction will remain a hot topic, at least for a decade. 



{\bf{Acknowledgments}}
The author is indebted to Rambahadur Gupta, Anu Sreedevi, and Bidisha Dey for their assistance in preparing several figures and collecting/cross-checking the predicted values of solar cycles 24 and 25. Ram and Anu helped me considerably in extracting several references for solar cycle predictions. I am also grateful to Arnab Rai Choudhuri (my PhD supervisor), Jie Jiang, Piyali Chatterjee, and Dibyendu Nandi, from whom I have learned most of the fundamentals of the solar dynamo during my PhD, which enriched my understanding of the solar cycle and its prediction, as reflected in this review. Now I acknowledge financial support provided 
the Indian Space Research Organisation (project no. ISRO/RES/RAC-S/IITBHU/2024-25) and the Anusandhan National Research Foundation (ANRF) through the MATRIC program (file no. MTR/2023/000670)
and the computational resources of the PARAM Shivay Facility under the National Supercomputing Mission, the Government of India, at the Indian Institute of Technology Varanasi. SOHO is a project of international cooperation between ESA and NASA. Courtesy of NASA/SDO and the HMI science teams. The group sunspot number data are obtained from SIDC/SILSO (https://www.sidc.be/SILSO/DATA/GroupNumber/) and the sunspot number data are from World Data Center (WDC)-SILSO, Royal Observatory of Belgium, Brussels, International Sunspot Number V2.0 \citep{SILSO_Sunspot_Number}.  
The ``Flare Index" dataset was prepared by the Kandilli Observatory and Earthquake Research Institute at the Bogazici University and made available through the NOAA National Geophysical Data Center (NGDC;  https://www.ngdc.noaa.gov/stp/space-weather/solar-data/solar-features/solar-flares/index/)

\appendix

%


\begin{longtable}{l|l|l}
        \caption{Summary of the predictions for Cycle 25.}
     \\ \hline
     \bf{Publications}    & \bf{Predicted} & \bf{Method} \\
     & \bf{sunspot} &  \\
     & \bf{number} &  \\
     \hline
     \citet{DuNDu2006b}  & $111.6 \pm 17.4$ & Descending time of the previous cycle \\
     \citet{Hamid2006} & $90.7\pm9.2$ & Counts and durations of spotless events \\
     \citet{Kane2007b} & $119 \pm 7.5$  & Extrapolation of spectral components \\
     \citet{Abdusamatov2007} & $50\pm15$ & Variations in solar radius or solar constant\\
     \citet{Hiremath2008} & $110 \pm 11$ & Physics-based (oscillator) model\\
     \citet{Quassim2007} & 116 & Neuro-fuzzy approach \\
     \citet{Pishkalo08}  &  $112.3 \pm 33.4$ & Neural-network-based sunspot prediction \\
     \citet{Rigozo2011} & $132.1$ & Extrapolation of spectral components \\
     \citet{HELAL2013} & 118.2 & Statistical method of precursors\\
     \citet{Yoshida14} & $115.4 \pm 11.9$ & Sunspot number before the minimum \\
     \citet{Javaraiah2015} & $50\pm15$ & Empirical correlations of cycle parameters\\
     \citet{Li2015} & 109.1 & Multi-parameter empirical extrapolation\\  
     \citet{bisoi} & $56 \pm 12$ & Declining of solar \& interplanetary fields\\
     \citet{Cameron2016} & Moderate & Physics based model (SFT) \\
     \citet{Gkana16} & 54.4 & Time series analysis of sunspot number \\
\citet{Dani2019}    & $59.4 \pm 22.3$ & Machine Learning Regression methods \\  
\citet{Kitiashvili2016} & $90\pm15$ & Data assimilation with Kalman filter\\
\citet{HU16} & $98.9 \pm 42$ & Data driven SFT model \\
\citet{Singh2017} & $102.8 \pm 24.6 $ & Using Hurst exponent \\
\citet{Chae17} & 153 & Autoregressive model\\
\citet{Iijima17} & $<$ Cycle 24 & Dipole moment from SFT model\\
\citet{Podladchikova} & $<$ Cycle 24  & Sunspot number before minimum\\
\citet{Kakad2017} & 63 & Shannon Entropy-Based Prediction\\
\citet{UH18} & $107 \pm 17$ & Data-driven SFT simulations\\ 
\citet{Hawkes2018} & 117 & Using Magnetic Helicity \\
\citet{BN18} & $107.63\pm17$ & Data-driven SFT simulations\\
\citet{Okoh2018}  & $122.1 \pm 18.2$ & Hybrid Regression-Neural Network \\
\citet{jiang2018predictability} & $124\pm 31$ &  Data-driven SFT simulations\\
\citet{Attia2013} &$ 90.7\pm8$ & Regression and correlation-based\\
\citet{Petrovay2018} & 129 & Rush-to-the-poles in coronal emission\\
\citet{Pesnell2018} & $135\pm25$ & Polar  field precursor\\
\citet{Gopalswamy2018} &$148 \pm 14$ & Proxy of polar field \\
\citet{Sarp2018} & $154\pm12$ & Nonlinear time-series modeling\\
\citet{Li2018} & $168.5 \pm 16.3$ &  Solar cycle characteristics\\
\citet{Sabarinath18} & $85\pm10$ & Multivariate regression \& binary mixture\\
\citet{Hawkes2018} & 117.3 & Magnetic Helicity as a Precursor \\
\citet{Okoh2018} & $124 \pm 20.7$ & Hybrid Regression-Neural Network \\
\citet{Macario-Rojas} & $96.98 \pm 22.09$  & Dynamo model with observed data\\
\citet{Kirov18} & 78 & Geomagnetic activity as a precursor\\
\citet{Covas2019} & $57\pm 17$ & Neural Network forecast\\
\citet{Labonville2019} & $105^{+29}_{-14}$ & Dynamo with data assimilation\\
\citet{Dani2019} & $159.4\pm 22.3$ & Machine Learning based model\\
\citet{sello2019} & $107\pm10$ & Non-linear dynamic techniques \\
\citet{Dani2019} & $159.4 \pm 22.3$ & Frequency-domain analysis of sunspot\\
\citet{HC19} & $139.34 \pm 16$ & Polar field and decay rate of cycle\\
\citet{Han2019} & $228.8 \pm 40.5$ & Neural-network using solar indices\\
\citet{Dani2019} & $110.2 \pm 12.8$ & ML-Regression methods\\
\citet{Singh19} & $89 \pm 9$ &  Hodrick Prescott filtering method\\
\citet{Pala19} & $163.7\pm10$ &  Deep Learning Methods \\
\citet{Tan19} & $<$ Cycle 24 & Statistical investigation of past cycles\\
\citet{Kakad20} & $136.9 \pm 24$ & Analysis using kernel density estimator\\
\citet{Uptonpanel19} & $<$ average & Survey of forecasts by Panel\\
\citet{McIntosh20} & $>$ Cycle 23 & Extended solar cycle model\\
\citet{Svalgaard20} & $128 \pm 
 10$ & Polar field at minimum\\
\citet{Lee20} & 101.44 & Hybrid Deep Learning Model\\
\citet{Benson20} & $107.2 \pm 19.75$ & Deep Neural Networks\\
\citet{Goncalves} & 117 & Warped Gaussian process regression\\
\citet{Miao20} & $121.5\pm32.9$ & Precursor Method (aa-index)\\
\citet{bisoi20} & $134\pm11$  &  Polar  and heliospheric field at 1 AU\\
\citet{Bisoi20b} & $155 \pm 11$ & Flux cancelled at equator \& polar field\\
\citet{du20} & $137.8\pm31.3$& Comparing the past similar cycles\\
\citet{Du20a} & $130\pm31.9$ & Sunspot number before solar minimum\\
\citet{Du20b} & $151.1\pm16.9$ & Waldmeier effect\\
\citet{Kumar21} & $120 \pm 25$ & Polar field rise rate \\
\citet{Guo21} & 126 & Dynamo model with polar field\\
\citet{Coban21} & 59.4 & Deep learning technique\\ 
\citet{Javaraiah21} & $76 \pm 12$ & Properties of solar cycle\\
\citet{Velasco21} & $95\pm15$ & Machine Learning algorithms\\
\citet{Courtillot21} & $97.6 \pm 7.8$ & Astronomical ephemeris\\
\citet{Chowdhury21} & $100.21\pm 15.06$& Geomagnetic activity Ap-index\\
\citet{Burud21} & $104.23\pm17.35$ & Spotless days and geomagnetic index\\
\citet{Wu21} & $114.5\pm16$&  Two-parameter modified logistic\\
\citet{Singh21} & $119.42\pm23.51$& Geomagnetic activity indices\\
\citet{Janssens21} & $118\pm29$ & Polar faculae\\
\citet{Xiong21} & 140.2 & Several precursors\\
\citet{Yan21} & $120\pm27$ & Two-stage Statistical Model\\
\citet{Pishkalo21} & $121 \pm 4$ & Polar field near  minimum\\
\citet{Wang21} &114.3 & long short-term memory deep learning \\
\citet{Penza21} &$130\pm10$ & Relation of sunspot areas and plages\\
\citet{Li21} & 175.62 & Deep learning based on sunspot area\\
\citet{KK21} & $103.3\pm15$& Even-odd pair of solar cycle\\
\citet{Kumar22} & $137\pm23$ & Rise rate of polar field\\
\citet{Ahluwalia22} & $126.7 \pm 4.7$ & Sunspot numbers \& geomagnetic indices\\
\citet{Bizzarri22} & $110\pm9$ & Multistep Bayesian Neural network \\
\citet{Brajva22} & $121\pm33$ & Activity level 3 yr before minimum\\
\citet{Du22} & $124\pm30$& Fitting the solar cycle\\
\citet{Javaraiah22} & $130\pm12$ & Long-Term Variations in Solar Activity\\
\cite{Dang22} & $133.47$ & Various ML models\\ 
\citet{Du22b} & $135.5\pm33.2$ & Rise rate\\
\citet{Zhu22} & $147.9$ & Optimized Long Short-Term Memory\\ 
\citet{Prasad22} & $171.9\pm3.4$ & Long short-term memory forecasting\\
\citet{Efimenko22} & $184\pm15$ & Rate of increase of solar activity\\
\citet{Lu22} & 145.3 &  Bimodal forecasting model\\
\citet{Nagovitsyn22} & 147 & Gnevyshev-Ohl rule\\
\citet{Ivanov22} & $181\pm46$ & Cycle length vs amplitude relation\\
\citet{Chowdhury22} & $160.29 \pm 66.06$ & Non-linear empirical dynamical modelling\\
\citet{Podladchikova22} & $110\pm26$ & Growth rate of the cycle\\
\citet{Arregui22} & $190.6\pm10$ & Bayesian inference\\
\citet{Koutchmy22} & 226.6 & Polar regions activity\\
\citet{Espuna-Fontcuberta23} & $134\pm8$ & Neural--network based ML technique\\
\citet{Obridko23} & $125.2\pm5.6$ & Extended solar cycle model\\
\citet{Javaraiah23} & $125\pm11$ & Variations of sunspot area and number\\
\citet{UH23} & $134 \pm 8$ & Curve-fitting and precursors \\
\citet{Prasad23} & 136.9 & Deep Learning Based Neural Network\\
\citet{Nagovitsyn23} & $149\pm28$ & Sunspot number 7 yr before minimum\\
\citet{Becheker23} & $164\pm10$ & Curve fitting \\
\citet{zhu23}  & $214.4\pm22$ & Long short-term memory model \\
\citet{Liu23}  & $135.2$ & Long short-term memory model \\
\citet{Raju23} & $206 \pm 7.6$
& Supergranular lane and sunspot number\\
\citet{Aparicio23} & $131\pm32$ & Ascending inflection point \\
\citet{Pishkalo23} & $136 \pm 36$ & Cycle precursor \\
\citet{Efimenko23} & $185\pm18$ &
Sunspot number increase rate
\\
\citet{Kalkan23} & 116.6 & Neural networks model\\
\citet{Arregui23} & $184\pm25$ & Bayesian Interface\\
\citet{Riley23} & 130 &  Superposed-epoch analysis\\
\citet{Su23} & $133.9 \pm 7.2$ & Deep learning model\\
\citet{Lozitsky23} & $183 \pm 23$ & Rise rate of sunspot number\\
\citet{Su24} & $127.94\pm6.37$ & Deep learning models \\
\citet{Flandez24} & 179 & Visibility Graph and Hathaway function\\
\citet{Sharma24} & $112\pm18$ & Geomagnetic precursor\\
\citet{Rodr24} & $134.2\pm9.8$ & Multivariate ML\\
\citet{He24} & 160 & ML model\\
\citet{Kumarvipin24} & 
$127.16\pm6.83$& Deep learning model\\
\citet{Petrovay24} & $155.8 \pm20.7$ & Parametric time series model\\
\citet{KK23} & 140.84 & ML model\\
\citet{Rodr24b} & $67.75\pm6.45$ & Spectral Analysis \& ML model\\
\citet{Cao24} & $141.55\pm36.95$ & ML model\\
\citet{PN24} &$128\pm20$ & Algebraic quantification of dipole moment\\
\citet{Luo24} & $146.7 \pm 33.40$ & Wavelet transform and fitting function\\
\citet{Xu24} & $193\pm39$ & Physics informed ML\\
\citet{Pesnell24} & $130.7\pm0.5$ & Slope of rising phase\\
\citet{Jaswal24} & $116.91\pm2.89$ & Dipole moment decay rate\\
\citet{Wang24} & $146\pm21$ & Interplanetary field strength\\
\citet{Kleeorin24} & 150  & Dynamo model \& neural network\\
  %
\label{tab:cyc25}
\end{longtable}

{\bf{Conflict of interest}}
The author states that there is no conflict of interest.

\bibliography{paper}

\end{document}